%% file: main.tex
\documentclass[aps,prb,reprint,10pt,amsmath,amssymb,superscriptaddress,longbibliography,floatfix]{revtex4-2}
\usepackage{placeins}
\usepackage[T1]{fontenc}
\usepackage{lmodern}
\usepackage{graphicx}
\usepackage{bm}
\usepackage{booktabs}
\usepackage{hyperref}
\hypersetup{colorlinks=true,linkcolor=blue,citecolor=blue,urlcolor=blue}
\graphicspath{{figures/}}
\newcommand{\vect}[1]{\bm{#1}}
\newcommand{\dd}{\mathrm d}
\newcommand{\BZ}{\mathrm{BZ}}
\newcommand{\BCD}{\mathrm{BCD}}
\newcommand{\kbT}{k_{\mathrm B}T}
\newcommand{\eV}{\,\mathrm{eV}}
\newcommand{\unit}{\mathrm{I}_2}
\newcommand{\av}[1]{\left\langle #1\right\rangle_{\vect k}}

\newcommand{\manuscripttitle}{Symmetry-Guided Machine Learning of Spin-Dependent Nonlinear Hall Response}
\newcommand{\manuscriptauthors}{Jyotirmoy Sau}
\newcommand{\manuscriptaffiliations}{Department of Physics and Astronomy, Uppsala University, Box-516, S-75120 Uppsala, Sweden}
\newcommand{\manuscriptemail}{jyotirmoy.sau@physics.uu.se}
\newcommand{\manuscriptacknowledgments}{J. S. acknowledges support from Uppsala University.}

\hypersetup{pdftitle={Symmetry-Guided Machine Learning of Spin-Dependent Nonlinear Hall Response},pdfauthor={\manuscriptauthors}}

\begin{document}
\makeatletter
\let\auto@bib@innerbib\@empty
\let\auto@bib\@empty
\makeatother
\let\originaladdcontentsline\addcontentsline
\renewcommand{\addcontentsline}[3]{}
\title{\manuscripttitle}
\author{\manuscriptauthors}
\email{\manuscriptemail}
\affiliation{\manuscriptaffiliations}
\begin{abstract}
A Berry-curvature dipole can generate a current perpendicular to an applied electric field even when the linear Hall response is zero. This work examines how lattice symmetry, magnetic orientation, and carrier occupation control this nonlinear response in an anisotropic honeycomb model. Unequal hopping permits one dipole component without magnetism. Rotating the local moments can activate the second component and change the sign and magnitude of both. Reversing all moments leaves the dipole unchanged but reverses the intrinsic linear Hall conductivity. Equal populations of reversed domains can therefore retain the dipole contribution when their local electric fields are the same. A symmetry-constrained polynomial machine-learning model describes the calculated dipoles for 1024 spin configurations. At the reference chemical potential, both components have $R^2>0.9995$ for 256 configurations excluded from training. Keeping the carrier density fixed shifts the chemical potential as the moments rotate and gives slightly larger prediction errors. Separate fits across a chemical-potential sweep show a stronger loss of accuracy for $D_y$ at higher energies. Within weak-field semiclassical transport with a constant relaxation time, the predicted dipole determines the angular dependence, rectification, and second harmonic of the current. The results connect magnetic control of the nonlinear Hall response to the balance of Berry-curvature contributions near the Fermi level.
\end{abstract}
\maketitle

\section{Introduction}\label{sec:introduction}
Berry curvature describes how an electronic wave function changes as its momentum moves through a crystal. It gives an electron an additional velocity perpendicular to an applied electric field and is central to the intrinsic Hall effect~\cite{Sundaram1999,Xiao2010}. The intrinsic linear Hall current depends on the sum of Berry curvature over occupied states. Positive and negative contributions can cancel, even when the curvature is large in individual regions of momentum space~\cite{PhysRevMaterials.9.024407,sau2024hydrostatic}. In a time-reversal-symmetric system, opposite momenta have opposite curvature, so the intrinsic linear Hall conductivity vanishes.

A nonlinear Hall current can survive this cancellation. An electric field changes both the electronic velocity and the distribution of occupied states. The product of these two changes produces a current proportional to the square of the field. One contribution is governed by the Berry-curvature dipole (BCD), which measures a directional imbalance of curvature near the Fermi surface~\cite{Moore2010,Sodemann2015}. Inversion symmetry forbids this dipole, while the remaining rotations and mirrors determine its allowed directions~\cite{Low2015,You2018}. Experiments on WTe$_2$ and corrugated bilayer graphene have demonstrated nonlinear Hall transport without breaking time-reversal symmetry~\cite{Ma2019,Kang2019,Ho2021}.

Symmetry determines whether a dipole is allowed, but it does not determine its size or sign. The dipole depends on Berry curvature, the direction of carrier motion, and which states lie near the Fermi level. Different bands and Fermi pockets can therefore add or cancel. Studies of Weyl semimetals, BiTeI, and bilayer WTe$_2$ show how nearby crossings and changes in band structure affect this balance~\cite{Zhang2018,Facio2018,Du2018}. Fermi-surface warping provides another route to a finite dipole~\cite{Battilomo2019}, while electrical control of Berry curvature illustrates the importance of carrier occupation~\cite{Xu2018}. A large local curvature alone is consequently insufficient to predict a large nonlinear Hall current.

Magnetic orientation provides an additional control. Spin--orbit coupling connects the moments to the crystal axes, so rotating them changes the band energies and wave functions. Noncollinear magnets can have Hall responses that depend on more than their net magnetization~\cite{Chen2014}. Symmetry-constrained spin expansions, combined with machine learning, have already been developed to describe such orientation-dependent Hall transport~\cite{Kipp2024}. The BCD is even under global moment reversal, whereas the intrinsic linear Hall conductivity is odd. Other rotations can still change the dipole's magnitude, direction, and sign, providing magnetic control of a response that is unchanged by complete reversal.

This work asks how to predict those changes in a model with independently specified lattice, magnetic, and carrier parameters. The calculation uses 1024 spin configurations of an anisotropic honeycomb model, together with nonmagnetic controls and globally reversed magnetic pairs. The controls separate the dipole permitted by the lattice from the changes caused by magnetic orientation. A symmetry-constrained polynomial model then predicts both dipole components for configurations excluded from training. The analysis focuses on the reversal-even BCD and how it depends on the carrier constraint.

Two carrier conditions are compared: a fixed chemical potential and a fixed number of electrons. They describe different responses because rotating the moments changes the band energies. At fixed electron number, the chemical potential must move to compensate, changing which states contribute to the dipole. Nonlinear Hall sign changes at fixed density have also been observed in graphene/black-phosphorus heterostructures~\cite{Ye2025}; here the control parameter is magnetic orientation. The comparison tests whether the same type of spin representation describes both carrier conditions and how its accuracy changes across an energy sweep.

The current calculated here is the BCD contribution in weak-field semiclassical transport with a constant relaxation time. Disorder can produce additional nonlinear Hall currents~\cite{Nandy2019,Du2019,Du2021,Duan2022}, and magnetic systems can also support intrinsic contributions associated with other aspects of quantum geometry~\cite{Liu2021,Gao2023,Wang2023}. The analysis below follows the BCD contribution from the band geometry to its angular and harmonic current signatures.

\section{Model and physical symmetry constraints}\label{sec:model}
The primitive cell contains sites $A$ and $B$, each with one orbital and a spin-$1/2$ electronic degree of freedom. Their prescribed local moments, $\vect s_A$ and $\vect s_B$, are classical unit vectors. In the basis $(A\uparrow,A\downarrow,B\uparrow,B\downarrow)$,
\begin{equation}
 H(\vect k)=\begin{pmatrix}
 \Delta\unit+J\vect s_A\cdot\vect\sigma&T(\vect k)\\
 T^\dagger(\vect k)&-\Delta\unit+J\vect s_B\cdot\vect\sigma
 \end{pmatrix},\label{eq:H}
\end{equation}
with nearest-neighbor hopping and Rashba coupling
\begin{equation}
 T(\vect k)=\sum_{j=1}^{3}e^{i\vect k\cdot\vect\delta_j}
 \left[-t_j\unit+i\alpha_R(\hat\delta_{jx}\sigma_y-\hat\delta_{jy}\sigma_x)\right].\label{eq:T}
\end{equation}
Here $a=|\vect\delta_j|$, $\hat{\vect\delta}_j=\vect\delta_j/a$, and $\vect\sigma$ contains the Pauli matrices. The $x$ bond has $t_1=1.25\eV$, while $t_2=t_3=1.00\eV$; $\alpha_R=0.40\eV$, $J=0.30\eV$, and $\Delta=0.50\eV$. The reference carrier conditions are $\mu_0=0.70\eV$ and $\kbT=0.05\eV$. The Supplemental Material (SM)~\cite{SupplementalMaterial}, Sec.~\ref{S-sec:geometry}, specifies the lattice vectors and Bloch convention.

\begin{figure}[t]
\includegraphics[width=\columnwidth]{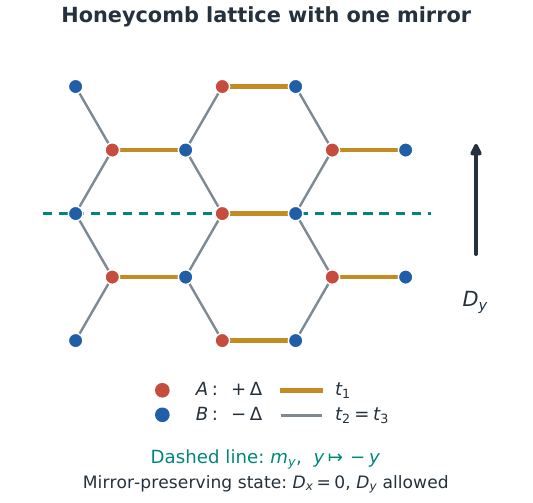}
\caption{Honeycomb model with inequivalent sublattices and a distinguished hopping direction. The structural mirror is the $xz$ plane. It allows $D_y$ and forbids $D_x$ when the full state preserves the mirror. }\label{fig:model}
\end{figure}

Figure~\ref{fig:model} identifies the two structural ingredients that permit the response. The red and blue sites carry opposite on-site potentials, breaking the inversion that exchanges them. The gold bonds distinguish the $x$ direction from the two equivalent oblique directions, removing threefold rotation while retaining the dashed mirror line $m_y:(x,y)\mapsto(x,-y)$. The $D_y$ arrow denotes the allowed dipole direction normal to that line; symmetry fixes this direction but does not determine a finite magnitude. Rashba coupling links spin orientation to the crystal axes, while exchange couples the prescribed moments to the electronic states. Rotating the moments can therefore change the response even though the bond geometry is fixed. Anisotropy is an electronic model parameter: no lattice deformation or next-nearest-neighbor hopping is introduced. The calculation treats a periodic two-site magnetic cell.

\begin{figure*}[t]
\includegraphics[width=0.75\textwidth]{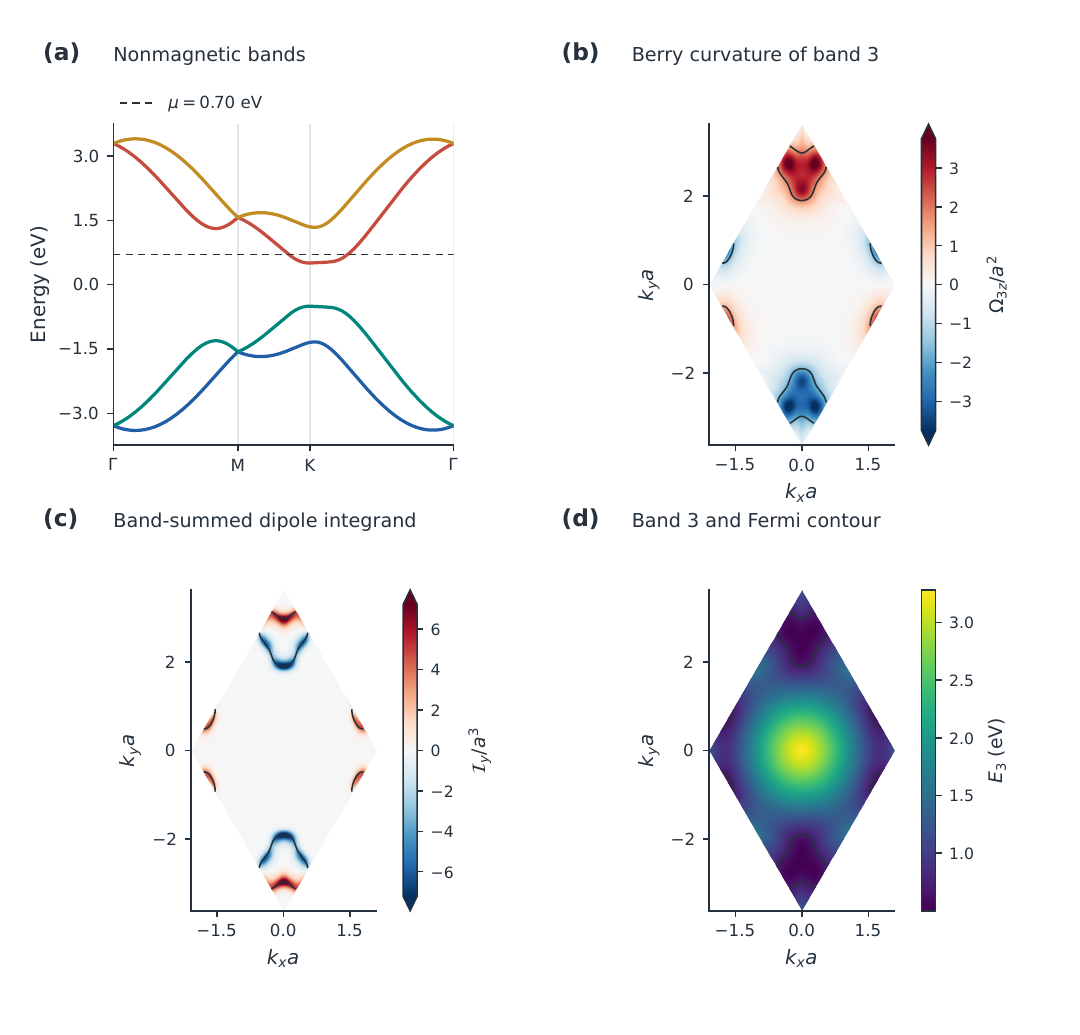}
\caption{Electronic geometry of the nonmagnetic $C_s$ control. (a) Bands along the geometric path $\Gamma$--$M$--$K$--$\Gamma$. (b) Curvature of the third energy-ordered band. (c) Full band-summed dipole integrand $\mathcal I_y=\sum_\nu\Omega_{\nu z}(\partial_{k_y}E_\nu)(-\partial_E f_\nu)$. (d) Third-band energy. The black contour marks $E_3=\mu_0$.}\label{fig:bands}
\end{figure*}

The dipole is
\begin{align}
 D_a&=\sum_\nu\int_{\BZ}\frac{\dd^2k}{(2\pi)^2}
 f_\nu\,\partial_{k_a}\Omega_{\nu z}\nonumber\\
 &=\sum_\nu\int_{\BZ}\frac{\dd^2k}{(2\pi)^2}
 \Omega_{\nu z}\,\partial_{k_a}E_\nu\,(-\partial_{E_\nu}f_\nu),\label{eq:D}
\end{align}
where $a=x,y$, $f_\nu$ is the Fermi function, and all four bands are included. The second form follows by integration by parts and makes the carrier selection explicit~\cite{Sodemann2015}. The factor $-\partial_Ef$ selects an energy window around $\mu$, while $\partial_{k_a}E=\hbar v_a$ assigns a sign according to the direction of motion. The dipole has dimensions of length; results are reported as $D_a/a$. The electronic temperature broadens this occupation window. Normalization and the Fermi-surface interpretation are developed in SM Sec.~\ref{S-sec:transport}.

Two cancellation arguments explain why inversion and time reversal act differently. If the full system has inversion symmetry, curvature is even in momentum and velocity is odd, making the dipole integrand odd. The dipole then vanishes. In a time-reversal-symmetric system, both curvature and velocity are odd, and their product is even. A finite dipole can therefore coexist with zero intrinsic linear Hall conductivity. Broken inversion is necessary for this channel but does not guarantee a nonzero result: additional rotations or mirrors can still enforce cancellation.

For a mirror-preserving state of the present model, $\Omega_z(k_x,-k_y)=-\Omega_z(k_x,k_y)$. The $x$ velocity is even under the reflection, so the $D_x$ integrand is odd and cancels. The $y$ velocity is odd, so the $D_y$ integrand is even and can survive. The permitted dipole is consequently perpendicular to the mirror line. This is the single-mirror selection familiar from two-dimensional BCD theory~\cite{Sodemann2015,You2018}. Restoring threefold symmetry in the nonmagnetic model removes any invariant in-plane dipole direction.

For arbitrary spin pairs $\vect q=(\vect s_A,\vect s_B)$, the corresponding covariance relations are
\begin{align}
 \vect D(-\vect q)&=\vect D(\vect q),
 &\sigma_{xy}(-\vect q)&=-\sigma_{xy}(\vect q),\label{eq:TR}\\
 \vect D(m_y\vect q)&=(-D_x(\vect q),D_y(\vect q)).\label{eq:mirror}
\end{align}
Spins are axial vectors: the mirror sends $(s_x,s_y,s_z)$ to $(-s_x,s_y,-s_z)$. These equations compare related configurations; neither time reversal nor the mirror need be a symmetry of one generic magnetic state. A mirror-breaking orientation can therefore activate $D_x$ while retaining the globally even spin-reversal parity of both components. SM Secs.~\ref{S-sec:cancellations} and \ref{S-sec:expansion} give the derivations. The allowed components are thus fixed by symmetry, while their magnitude and sign depend on the momentum-space contributions examined next.

\begin{figure*}[t]
\includegraphics[width=0.95\textwidth]{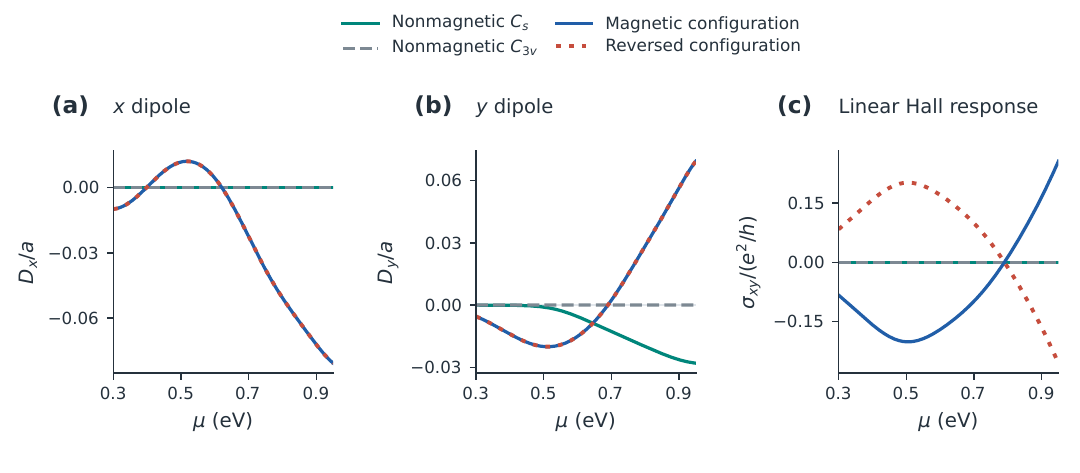}
\caption{Direct symmetry controls: (a) $D_x$, (b) $D_y$, and (c) intrinsic linear Hall conductivity. The nonmagnetic $C_s$ model permits $D_y$, whereas $C_{3v}$ symmetry removes both components. The corresponding selection rules are summarized in Table~\ref{tab:symmetrycontrols}. A magnetic configuration (sample ID 0) and its globally reversed partner have identical dipoles and opposite linear Hall conductivities. Structural cancellation and magnetic reversal therefore test different properties of the response.}\label{fig:controls}
\end{figure*}

\section{Curvature cancellation and magnetic control}\label{sec:controls}
The relation between a curvature map and a transport coefficient is clearest in the zero-temperature limit. For regular Fermi contours, Eq.~\eqref{eq:D} becomes
\begin{equation}
 D_a=\sum_\nu\oint_{E_\nu=\mu}\frac{\dd\ell}{(2\pi)^2}
 \Omega_{\nu z}\frac{\partial_{k_a}E_\nu}{|\nabla_{\vect k}E_\nu|}.\label{eq:contour}
\end{equation}
The ratio is a component of the contour normal directed toward increasing energy. Opposite portions of a pocket therefore contribute with different signs even if their curvatures have the same sign. The BCD measures the imbalance that remains after summing this oriented curvature over all pockets and bands. It is not a quantized topological invariant. At finite temperature the result is an energy average of these contour contributions, as derived in SM Sec.~\ref{S-sec:spectral}.

Avoided crossings provide one route to a large local contribution because Berry curvature contains interband matrix elements divided by squared band separations. A small separation can enhance curvature where eigenstates mix rapidly, consistent with the band-geometry mechanisms identified in related systems~\cite{Facio2018,Du2018}. Nevertheless, the transport contribution can remain weak if the feature lies outside the occupation window, has an unfavorable velocity projection, or is compensated by another band. This is why the band-summed dipole integrand is essential to the interpretation.

Setting $J=0$ removes exchange and isolates the structural contribution. Figure~\ref{fig:bands}(a) places the reference chemical potential relative to the four bands along a geometric reciprocal-space path. The energy map in Fig.~\ref{fig:bands}(d) complements this one-dimensional cut by locating the third-band Fermi contour throughout the displayed reciprocal cell. That contour is also drawn on the curvature map in Fig.~\ref{fig:bands}(b): the opposite colors at opposite momenta display the time-reversal cancellation of the equilibrium Hall response. Their velocities also reverse, making their curvature--velocity products equal. The field-sensitive weighting can therefore reinforce contributions that cancel in linear response.

Comparing Figs.~\ref{fig:bands}(b) and \ref{fig:bands}(c) shows how curvature becomes a transport contribution. Panel (c) includes all four bands, weighted by the velocity projection and the finite-temperature occupation window. Its positive and negative regions therefore need not coincide with the largest curvature in panel (b). The third-band Fermi contour in panel (d) helps locate the contributing states but represents only part of the four-band sum. Integrating the full signed integrand at the reference carrier conditions gives
\begin{equation}
 D_y/a=-0.0127421,\qquad D_x=0,\qquad\sigma_{xy}=0.\label{eq:nonmagnetic}
\end{equation}
The last two equalities are symmetry-enforced zeros recovered numerically within $5\times10^{-16}a$ and $8\times10^{-17}e^2/h$. The negative $D_y$ means that negative signed curvature--velocity contributions dominate the integral in the adopted convention; it does not mean that the curvature is negative everywhere.

The equal-hopping, nonmagnetic control restores $C_{3v}$ symmetry. Contributions from three rotation-related regions then sum to zero because an in-plane dipole cannot remain invariant under a $120^\circ$ rotation. Figures~\ref{fig:controls}(a) and \ref{fig:controls}(b) verify this cancellation over the displayed chemical-potential interval, with a residual no larger than $2.03\times10^{-9}a$. Comparing the two nonmagnetic curves separates the restrictions: both have zero $D_x$ in panel (a), but only the single-mirror $C_s$ control retains $D_y$ in panel (b). Their linear Hall conductivities remain zero in panel (c). Unequal hopping therefore removes a specific rotational cancellation without requiring a magnetic moment to create the nonlinear response. Equal hopping would not enforce a zero dipole for arbitrary moments that themselves break the rotation.

\begin{figure*}[t]
\includegraphics[width=0.95\textwidth]{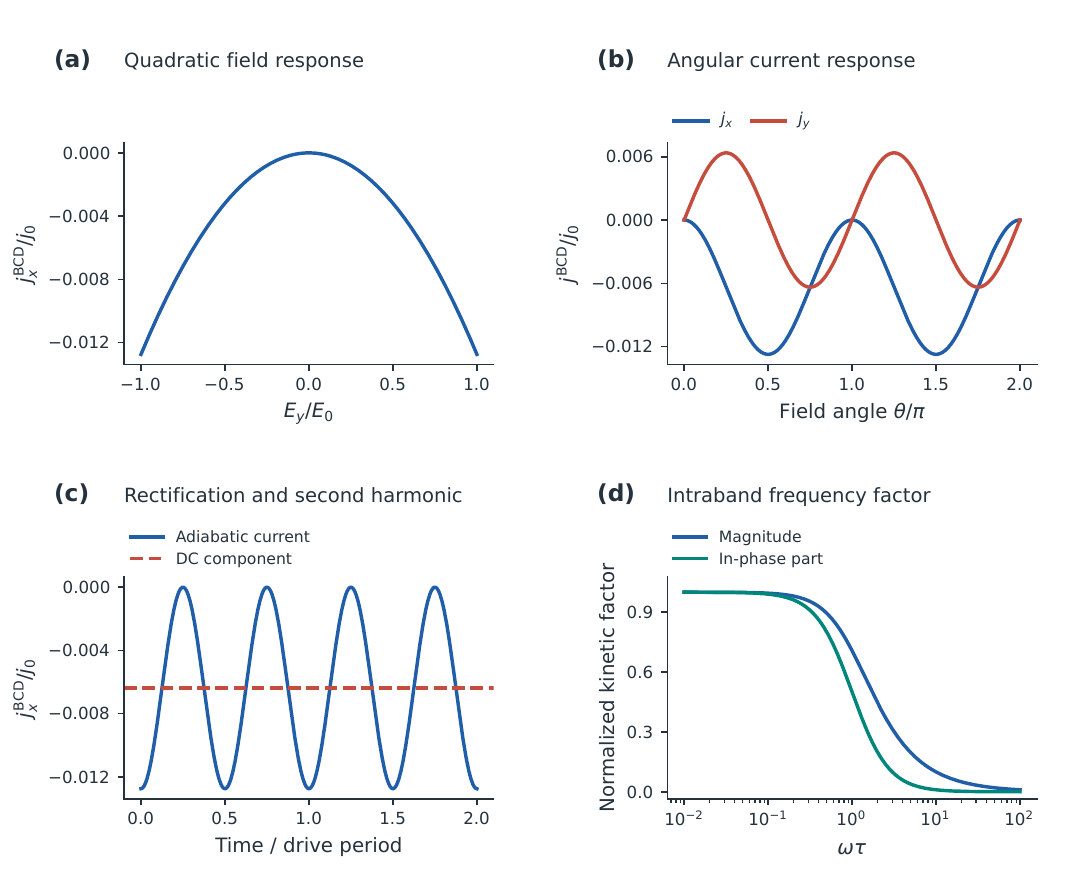}
\caption{Electrical consequences of the calculated nonmagnetic dipole: (a) quadratic field dependence, (b) angular response, (c) slow-drive rectification and second harmonic, and (d) the constant-$\tau$ frequency factor. Currents are normalized by $j_0=e^3\tau aE_0^2/\hbar^2$. Panels (a)--(c) follow from the calculated dipole and the weak-field current formula. Panel (d) describes intraband relaxation.}\label{fig:current}
\end{figure*}

\begin{table*}[t]
\caption{Symmetry constraints underlying Fig.~\ref{fig:controls}. The nonmagnetic $C_s$ dipole value is evaluated at $\mu_0=0.70\eV$; zeros denote symmetry constraints. Global moment reversal compares $\vect q=(\vect s_A,\vect s_B)$ with $-\vect q$.}\label{tab:symmetrycontrols}
\centering
\renewcommand{\arraystretch}{1.15}
\begin{tabular*}{0.95\textwidth}{@{\extracolsep{\fill}}lll@{}}
\toprule
Control or transformation & Berry-curvature dipole & Intrinsic linear Hall conductivity\\
\midrule
Nonmagnetic $C_s$ & $D_x=0,\quad D_y/a=-0.0127421$ & $\sigma_{xy}=0$\\
Restore nonmagnetic $C_{3v}$ & $D_x=D_y=0$ & $\sigma_{xy}=0$\\
Reverse both moments & $\vect D(-\vect q)=\vect D(\vect q)$ & $\sigma_{xy}(-\vect q)=-\sigma_{xy}(\vect q)$\\
\bottomrule
\end{tabular*}
\end{table*}

Exchange changes both the eigenstates and their energies. Generic moment orientations also remove the mirror, releasing the cancellation of $D_x$. In Figs.~\ref{fig:controls}(a) and \ref{fig:controls}(b), the magnetic configuration and its reversed partner trace the same two-component dipole curves, including their zero crossings. In Fig.~\ref{fig:controls}(c), the same reversal instead reflects the linear Hall curve about zero. The dipole retains its sign because both curvature and velocity reverse when the two configurations are compared at opposite momenta. Changes of sign with chemical potential express a change in the balance of signed contributions; they do not by themselves establish a magnetic or topological transition. Strong orientation dependence is therefore fully compatible with even parity under complete reversal.

Table~\ref{tab:symmetrycontrols} summarizes the structural selection rules and the contrasting reversal parities of the dipole and intrinsic linear Hall conductivity.

An immediate consequence concerns domain averaging. Consider only the bulk responses of domains $\vect q$ and $-\vect q$, with common local electric field and relaxation time. If their fractions are $p$ and $1-p$, then
\begin{equation}
 \overline{\vect D}=\vect D(\vect q),\qquad
 \overline{\sigma}_{xy}=(2p-1)\sigma_{xy}(\vect q).\label{eq:domains}
\end{equation}
Equal populations cancel the linear Hall signal while retaining the BCD contribution. Thus this channel can be insensitive to reversal-domain balance while remaining sensitive to other moment rotations. It cannot distinguish the two reversed domains individually. Equation~\eqref{eq:domains} is a symmetry-derived implication, not a simulation of domain walls or current redistribution. SM Sec.~\ref{S-sec:domains} states the averaging assumptions and the different cancellation for mirror-related domains.

These conclusions require resolving small differences between large signed momentum-space contributions. Narrow avoided crossings can make even the sign from a uniform grid unreliable. SM Fig.~\ref{S-fig:convergence} demonstrates this risk and the stability recovered by adaptive integration. The final orientation dataset uses that refinement, tighter-tolerance audits, and agreement between the two forms of Eq.~\eqref{eq:D}; the full evidence is in SM Sec.~\ref{S-sec:numerics}. The dipole obtained from this integration determines the angular and harmonic current response discussed below.

\section{Rectification and angular response}\label{sec:current}
The quadratic current combines the anomalous velocity with the field-induced occupation shift. Both are first order in the electric field, and the occupation shift is proportional to the momentum relaxation time. Their product gives the curvature--velocity weighting in Eq.~\eqref{eq:D}. For electron charge $-e$ ($e>0$), Berry connection $i\langle u|\nabla_{\vect k}u\rangle$, and constant relaxation time $\tau$, the static BCD sheet current is
\begin{equation}
 \vect j^{\BCD}=\frac{e^3\tau}{\hbar^2}
 (\vect E\times\hat{\vect z})(\vect D\cdot\vect E).\label{eq:j}
\end{equation}

The prefactor contains scattering through $\tau$, even though the dipole itself is a band-geometric quantity~\cite{Sodemann2015}. SM Sec.~\ref{S-sec:boltzmann} derives the sign and prefactor from the occupation correction.
The band-resolved semiclassical description applies when the electronic states contributing appreciably to transport remain
well resolved on the scales set by scattering and the driving frequency, and the electric field is sufficiently weak to suppress
interband mixing. For a relevant local band separation $g$, this corresponds to scattering linewidths and photon energies
small compared with $g$. The electronic temperature controls the occupation weights independently of these dynamical scales.
The frequency dependence considered below describes intraband relaxation within this regime.

Equation~\eqref{eq:j} separates two roles of geometry. The projection $\vect D\cdot\vect E$ selects the field directions that activate the response, whereas $\vect E\times\hat{\vect z}$ fixes its transverse direction. The response vanishes for $\vect E\perp\vect D$ and has its largest magnitude at fixed field strength for $\vect E\parallel\vect D$. Reversing the field reverses both factors, leaving the current unchanged. Conversely, reversing the dipole through a change of magnetic orientation reverses the quadratic current at fixed field.

Figure~\ref{fig:current}(a) converts the nonmagnetic control into a current for a field along $y$:
\begin{equation}
 \frac{j_x^{\BCD}}{j_0}=-0.0127421\left(\frac{E_y}{E_0}\right)^2,
 \qquad j_0=\frac{e^3\tau a E_0^2}{\hbar^2}.\label{eq:currentnumber}
\end{equation}
The downward parabola makes the negative dipole and even field parity visible: either sign of $E_y$ drives a current along $-x$. For a rotating field $\vect E=E_0(\cos\theta,\sin\theta)$, the general angular response is
\begin{align}
 j_x/j_0&=\sin\theta\,(d_x\cos\theta+d_y\sin\theta),\nonumber\\
 j_y/j_0&=-\cos\theta\,(d_x\cos\theta+d_y\sin\theta),\label{eq:angle}
\end{align}
where $d_a=D_a/a$. In the mirror-preserving control, $d_x=0$, giving $j_x/j_0=d_y\sin^2\theta$ and $j_y/j_0=-d_y\sin\theta\cos\theta$. These are the two curves in Fig.~\ref{fig:current}(b). Both vanish for a field along $x$, perpendicular to the dipole; the total current magnitude is greatest for a field along $y$. Between these directions, the changing Cartesian components describe a current that remains perpendicular to the rotating field. A magnetic orientation with nonzero $D_x$ changes the inactive field direction. Angular measurements therefore access the direction of the dipole as well as its magnitude, although extracting a BCD contribution from a total experimental signal requires separating other mechanisms.

A slow drive $E_y(t)=E_0\cos\omega t$ produces
\begin{equation}
 \frac{j_x(t)}{j_0}=\frac{D_y}{2a}[1+\cos(2\omega t)].\label{eq:harmonic}
\end{equation}
Figure~\ref{fig:current}(c) shows the resulting two current oscillations per drive period, displaced from zero by the dashed dc level. Both field half-cycles drive the transverse current in the same direction, producing equal signed dc and second-harmonic amplitudes, approximately $-0.0063711j_0$ for the control. At finite frequency, the occupation shift acquires the kinetic factor $(1+i\omega\tau)^{-1}$: carriers respond less effectively when the field changes rapidly compared with their relaxation. Figure~\ref{fig:current}(d) distinguishes its magnitude from its in-phase part. The latter falls faster because the distribution also develops a phase lag, rather than merely a smaller response. These curves describe intraband relaxation; they do not calculate interband resonances. The harmonic conventions and frequency factors are derived in SM Sec.~\ref{S-sec:frequency}.

The identity $\vect j^{\BCD}\cdot\vect E=0$ means that this transverse contribution does no local electrical work by itself. Longitudinal transport and relaxation remain dissipative, however, and power conversion requires the full circuit response. Crystal-based rectification has been proposed through both skew scattering and nonlinear Hall transport~\cite{Isobe2020,Onishi2024}; the present result specifies a normalized BCD sheet current, not a device efficiency. An absolute current or Hall voltage requires a lattice scale, scattering information, longitudinal conductivity, and boundary conditions. SM Sec.~\ref{S-sec:frequency} gives the current conventions. Since both dipole components determine the angular response, the next section tests their prediction across magnetic orientations.

\begin{figure*}[t]
\includegraphics[width=0.95\textwidth]{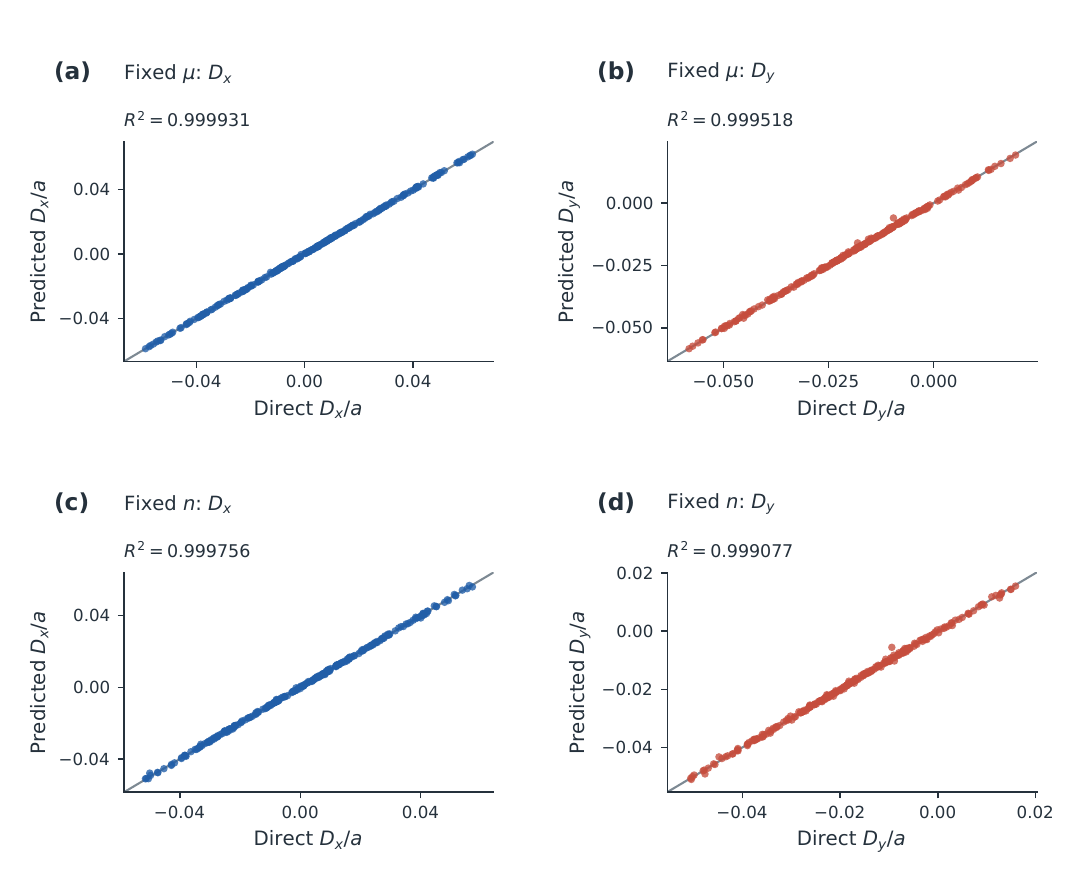}
\caption{Both dipole components predicted for the 256 held-out configurations using the larger feature budget: (a),(b) $D_x$ and $D_y$ at fixed chemical potential; (c),(d) $D_x$ and $D_y$ at fixed filling. The diagonal agreement establishes accurate interpolation of the sampled orientation dependence.}\label{fig:predictions}
\end{figure*}

\section{Spin dependence and carrier occupation}\label{sec:spin}
The calculation covers 1024 independently sampled spin pairs, uniform on the two unit spheres. These configurations explore orientation space without assigning an equilibrium magnetic probability. Of them, 768 determine the representation and 256 are excluded from fitting. The spin-to-dipole map uses polynomials of degrees zero, two, four, and six, constrained to have the mirror and reversal characters in Eqs.~\eqref{eq:TR} and \eqref{eq:mirror}. Sampling and statistical details are in SM Secs.~\ref{S-sec:sampling}--\ref{S-sec:features}; the main purpose is to test whether the electronic response can be predicted across different moment orientations.

The allowed spin dependence is physically richer than a dependence on net moment alone. Writing $\vect M=(\vect s_A+\vect s_B)/2$ and $\vect L=(\vect s_A-\vect s_B)/2$ separates the net and staggered coordinates. Both reverse under time reversal, so combinations of the forms $M_iM_j$, $L_iL_j$, and $M_iL_j$ can belong to the even response, subject to the component-specific mirror restriction. The staggered potential removes site-exchange symmetry, so mixed terms cannot be discarded on that basis. This decomposition explains what information an orientation-dependent response may contain; it does not establish the dominance of one term from a fit. SM Sec.~\ref{S-sec:magneticcoordinates} also discusses the unit-spin constraints and the limits of assigning a chirality mechanism to an individual coefficient.

At $\mu_0$, the sampled values span $-0.06586\le D_x/a\le0.06876$ and $-0.05888\le D_y/a\le0.01949$. Reorientation therefore changes both components' signs and magnitudes within the sample, even though global reversal leaves them unchanged. These are sampled ranges, not global extrema. The permitted constant in the $D_y$ representation reflects its mirror character; a finite-$J$ fitted offset is not the independently calculated $J=0$ dipole, because removing exchange changes the Hamiltonian.

Figures~\ref{fig:predictions}(a) and \ref{fig:predictions}(b) compare the directly calculated and predicted $D_x$ and $D_y$ at fixed chemical potential. Points follow the diagonal for both positive and negative dipoles, showing that the representation captures the magnitude and sign across untrained orientations, rather than only an average response. The larger representations give $R^2=0.999931$ for $D_x$ and $0.999518$ for $D_y$, with root-mean-square errors $2.4328\times10^{-4}a$ and $3.6642\times10^{-4}a$. Through Eq.~\eqref{eq:angle}, predicting both components specifies the orientation-dependent angular current pattern. SM Fig.~\ref{S-fig:fixedmu} resolves the remaining errors and their tails, which are less visible on a diagonal scatter plot.

These results show that symmetry-allowed orientation functions closely describe the response at the reference conditions. Higher polynomial degree provides finer angular dependence, but it is neither an expansion in small exchange nor evidence for a particular many-spin interaction. Unit-length spin constraints make the raw coefficients nonunique. The meaningful test is therefore the predicted dipole and its covariance, rather than an interpretation of individual coefficients as microscopic couplings.

The carrier constraint changes this function. At fixed $\mu_0$, the sampled filling varies from $1.057805$ to $1.083676$ electrons per site because moment rotation changes the band energies. Maintaining $n_0=1.065$ instead requires a configuration-dependent chemical potential between $0.624962$ and $0.728639\eV$. Figures~\ref{fig:predictions}(c) and \ref{fig:predictions}(d) show that both components remain close to the diagonal under this constraint. The fixed-filling fits yield $R^2=0.999756$ and $0.999077$ for $D_x$ and $D_y$, with root-mean-square errors $4.0092\times10^{-4}a$ and $4.5250\times10^{-4}a$. Both carrier conditions give accurate held-out predictions, although the absolute errors are larger at fixed filling for the chosen density, electronic temperature, and fitting procedure. SM Figs.~\ref{S-fig:residualfilling} and \ref{S-fig:fixedn} provide the filling-resolved residuals and the separate fixed-density test.

The physical origin of the distinction can be expressed without assuming a particular fitting method. Let $\lambda$ parameterize a smooth rotation of the moments. At fixed filling and electronic temperature,
\begin{equation}
 \left(\frac{\dd D_a}{\dd\lambda}\right)_n
 =\left(\frac{\partial D_a}{\partial\lambda}\right)_\mu
 -\left(\frac{\partial D_a}{\partial\mu}\right)_\lambda
 \frac{(\partial_\lambda n)_\mu}{(\partial_\mu n)_\lambda}.\label{eq:feedback}
\end{equation}
The first term describes moment rotation at fixed chemical potential. The second accounts for the chemical-potential shift needed to keep the filling constant and can reinforce or oppose the first term. Its denominator measures the change in filling with chemical potential. Fixed density therefore still allows the Fermi contours and their curvature weights to evolve. SM Sec.~\ref{S-sec:feedback} derives Eq.~\eqref{eq:feedback}. The numerical comparison evaluates the complete response under each carrier condition; it does not separately evaluate the two derivative contributions.

Sweeping $\mu$ moves the occupation window through different parts of the band structure. The finite-temperature dipole is a convolution of an energy-resolved curvature weight with $-\partial_Ef$ (SM Sec.~\ref{S-sec:spectral}), so the relative contributions of bands and Fermi contours change with energy. This motivates testing the representation across the sweep; a sign change or fit residual alone does not identify a phase transition.

Over $0.40$--$0.90\eV$, the larger-budget $D_x$ score remains between $0.998100$ and $0.999953$, whereas the $D_y$ score falls to $0.965260$ at the upper endpoint. There the $D_y$ root-mean-square error is $5.7499\times10^{-3}a$ and the largest held-out error is $4.7200\times10^{-2}a$. SM Fig.~\ref{S-fig:energy} shows how the orientation weights and representative dipole curves evolve as the occupation window moves. The deterioration persists after adaptive refinement of the transport integrals and therefore indicates reduced accuracy of the present surrogate at higher chemical potential. The available comparison does not separate the effects of polynomial truncation, feature selection, regularization, and training-set coverage. Each energy has separately fitted coefficients, so these results test orientation prediction at known energies, not extrapolation to an untrained energy. SM Sec.~\ref{S-sec:fits} gives the full comparison. 

\section{Discussion and conclusions}\label{sec:discussion}
The calculations show how lattice symmetry, magnetic orientation, and carrier occupation work together to control the Berry-curvature dipole. Unequal hopping removes the threefold-rotation constraint and allows a nonmagnetic $D_y$. Rotating the moments can break the remaining mirror and activate $D_x$. Both components can change sign because rotation changes the balance of curvature and velocity contributions near the Fermi level. Reversing all moments is different from a general rotation: it preserves the dipole while reversing the intrinsic linear Hall conductivity.

A symmetry-constrained polynomial machine-learning model captures this orientation dependence accurately at the reference chemical potential. For 256 held-out configurations, the reported scores are $R^2=0.999931$ for $D_x$ and $0.999518$ for $D_y$. These results describe interpolation between magnetic orientations of the same Hamiltonian. Keeping the carrier density fixed also gives accurate predictions, but with slightly larger errors. Maintaining the electron number requires the chemical potential to shift as the bands change, moving the energy window that selects the curvature contributions. Separate fits at other chemical potentials further show that the reference-energy accuracy is not maintained uniformly, especially for $D_y$ at the upper end of the sweep.

The dipole also gives clear current signatures within the stated transport approximation. The quadratic current is perpendicular to the electric field, vanishes when the field is perpendicular to the dipole, and is unchanged by electric-field reversal. A slow alternating field produces both a dc component and a second harmonic. Under a common local electric field and relaxation time, equally populated reversed magnetic domains retain this BCD current while canceling the intrinsic linear Hall current. These signatures follow from the BCD mechanism, but they do not by themselves exclude other nonlinear contributions~\cite{Nandy2019,Du2019,Liu2021}.

The model uses prescribed static moments in a periodic two-site cell and does not determine an equilibrium magnetic phase. Its main result is a quantitative connection between moment orientation and the two-component dipole, together with an explicit comparison of fixed chemical potential and fixed filling. This provides a basis for studying magnetic control of nonlinear Hall transport in more detailed electronic models, with the relevant magnetic states, carrier conditions, and numerical convergence checked for each system.

\begin{acknowledgments}
\manuscriptacknowledgments\par

\end{acknowledgments}

\nocite{Tsirkin2021,Zou2005,Pedregosa2011}

\input{main.bbl}
\clearpage
\onecolumngrid
\let\addcontentsline\originaladdcontentsline
\phantomsection\label{supplemental-material}
\input{supplemental.tex}
\end{document}

%% file: supplemental.tex

\setcounter{section}{0}
\setcounter{subsection}{0}
\setcounter{equation}{0}
\setcounter{figure}{0}
\setcounter{table}{0}
\setcounter{footnote}{0}
\setcounter{page}{1}
\renewcommand{\theHsection}{S\arabic{section}}
\renewcommand{\theHsubsection}{\theHsection.\Alph{subsection}}
\renewcommand{\theHequation}{S\arabic{equation}}
\renewcommand{\theHfigure}{S\arabic{figure}}
\renewcommand{\theHtable}{S\arabic{table}}
\renewcommand{\thesection}{S\arabic{section}}
\renewcommand{\thesubsection}{\thesection.\Alph{subsection}}
\makeatletter
\renewcommand{\p@subsection}{}
\makeatother
\renewcommand{\theequation}{S\arabic{equation}}
\renewcommand{\thefigure}{S\arabic{figure}}
\renewcommand{\thetable}{S\arabic{table}}
\renewcommand{\thepage}{S\arabic{page}}
\newcommand{\paperfigure}[4][165mm]{%
  \begin{figure}[!htbp]\centering
  \ifdim#1>\linewidth\def\smfigurewidth{\linewidth}\else\def\smfigurewidth{#1}\fi
  \includegraphics[width=\smfigurewidth,height=0.72\textheight,keepaspectratio]{figure_#2.pdf}
  \caption{#3}\label{S-fig:#4}
  \end{figure}}

\begin{center}
\bfseries Supplemental Material for ``\manuscripttitle''\par
\vspace{1em}
\normalfont\manuscriptauthors\par
\vspace{0.5em}
\small\itshape\manuscriptaffiliations\par
\normalfont\small\href{mailto:\manuscriptemail}{\manuscriptemail}\par
\end{center}

\tableofcontents

\clearpage
\section{Lattice, Hamiltonian, and parameters}\label{S-sec:geometry}
\label{S-sec:model}
The model consists of a periodic honeycomb lattice in the $xy$ plane with two sites, $A$ and $B$, per primitive cell. Each site carries one electronic orbital and a spin-$1/2$ degree of freedom. The prescribed local moments $\vect s_A$ and $\vect s_B$ are classical unit vectors. In the basis $(A\uparrow,A\downarrow,B\uparrow,B\downarrow)$, the Bloch Hamiltonian is
\begin{equation}
 H(\vect k)=\begin{pmatrix}
 \Delta\unit+J\vect s_A\cdot\vect\sigma&T(\vect k)\\
 T^\dagger(\vect k)&-\Delta\unit+J\vect s_B\cdot\vect\sigma
 \end{pmatrix},\label{S-eq:H}
\end{equation}
where $\vect\sigma$ is the vector of Pauli matrices and
\begin{equation}
 T(\vect k)=\sum_{j=1}^{3}e^{i\vect k\cdot\vect\delta_j}
 \left[-t_j\unit+i\alpha_R\left(\hat\delta_{jx}\sigma_y-\hat\delta_{jy}\sigma_x\right)\right].\label{S-eq:T}
\end{equation}
The nearest-neighbor distance is $a$, $\hat{\vect\delta}_j=\vect\delta_j/a$, and the bond vectors are
\begin{equation}
 \vect\delta_1=a(1,0),\qquad
 \vect\delta_2=a(-1/2,\sqrt3/2),\qquad
 \vect\delta_3=a(-1/2,-\sqrt3/2).\label{S-eq:bonds}
\end{equation}
The primitive translations are $\vect a_1=\vect\delta_1-\vect\delta_2$ and $\vect a_2=\vect\delta_1-\vect\delta_3$, giving $A_{\rm cell}=3\sqrt3 a^2/2$. Equations~\eqref{S-eq:H} and \eqref{S-eq:T} fix the embedded-position Bloch convention used for all derivatives. Momentum is expressed in units of $a^{-1}$ in the implementation, and the length unit is restored when reporting transport coefficients.

\begin{table}[!htbp]
\caption{Hamiltonian and sampling parameters. Energies are in eV. The electronic temperature smooths occupations; it is not a lifetime broadening or a temperature assigned to the sampled spins.}\label{S-tab:parameters}
\centering\small
\begin{tabular}{lll}\toprule
Quantity&Symbol&Value\\\midrule
Nearest-neighbor hopping, distinguished bond&$t_1$&$1.25$\\
Other nearest-neighbor hoppings&$t_2=t_3$&$1.00$\\
Rashba coupling&$\alpha_R$&$0.40$\\
Local exchange strength&$J$&$0.30$\\
Staggered onsite energy&$\Delta$&$0.50$\\
Electronic temperature&$\kbT$&$0.050$\\
Primary chemical potential&$\mu_0$&$0.70$\\
Chemical-potential sweep&$\mu$&$0.40,0.45,\ldots,0.90$\\
Fixed filling, electrons per site&$n_0$&$1.065$\\
Independent spin pairs&$N$&$1024$\\
Training / test configurations&&$768/256$\\
Maximum total polynomial degree&$p_{\max}$&$6$\\\bottomrule
\end{tabular}
\end{table}

The parameters are listed in table~\ref{S-tab:parameters}. The bond geometry is unchanged from an ideal honeycomb lattice. Anisotropy is introduced as an electronic hopping parameter; no elastic deformation is calculated. All hopping terms in \eqref{S-eq:T} connect nearest neighbors. The staggered term makes the two sites inequivalent, while $t_1\ne t_2=t_3$ removes threefold rotational symmetry. With Rashba coupling, the resulting nonmagnetic structural point group is $C_s$, generated by the mirror $m_y:(x,y)\mapsto(x,-y)$, shown in the \hyperref[fig:model]{main-text lattice figure}. This mirror preserves the two sublattices individually.

\FloatBarrier
\section{Symmetry and the spin-polynomial representation}\label{S-sec:symmetry}
A spin is an axial vector, so its transformation under this mirror is
\begin{equation}
 m_y:\ (s_x,s_y,s_z)\longmapsto(-s_x,s_y,-s_z).\label{S-eq:spinmirror}
\end{equation}
The out-of-plane Berry curvature is odd under this reflection. Its in-plane dipole therefore transforms as
\begin{equation}
 m_y:\ (D_x,D_y)\longmapsto(-D_x,D_y).\label{S-eq:Dmirror}
\end{equation}
These are covariance statements relating a magnetic configuration and its reflected partner. They do not imply $D_x=0$ for every configuration: that restriction applies only when the full magnetic state retains the mirror. A second control sets $t_1=t_2=t_3$ and $J=0$. The nonmagnetic point group is then $C_{3v}$, whose threefold rotation forces the in-plane dipole to vanish. Equal hopping alone does not impose this conclusion on an arbitrary magnetic state.

\subsection{Symmetry cancellations in the momentum integral}\label{S-sec:cancellations}
The dipole transforms as an in-plane axial vector in two dimensions: under an in-plane orthogonal operation $R$, $\vect D\mapsto(\det R)R\vect D$. This follows because $\Omega_z$ is a pseudoscalar under such operations while the momentum derivative is a polar vector. A reflection $R=\operatorname{diag}(1,-1)$ therefore gives Eq.~\eqref{S-eq:Dmirror}. For a state preserving this mirror,
\begin{align}
 E_\nu(k_x,-k_y)&=E_\nu(k_x,k_y),&
 \Omega_{\nu z}(k_x,-k_y)&=-\Omega_{\nu z}(k_x,k_y),\\
 \partial_{k_x}E_\nu(k_x,-k_y)&=\partial_{k_x}E_\nu(k_x,k_y),&
 \partial_{k_y}E_\nu(k_x,-k_y)&=-\partial_{k_y}E_\nu(k_x,k_y).
\end{align}
Band labels can be relabeled across the transformation; the all-band sum is the physical quantity. The occupation derivative is even under the mirror. Hence the $x$ dipole integrand is odd and integrates to zero, whereas the $y$ integrand is even. This is the single-mirror selection used in the main text~\cite{SM-Sodemann2015,SM-You2018}.

For inversion, the energy and axial curvature are even under $\vect k\mapsto-\vect k$, but the energy derivative is odd. Thus an inversion-symmetric full state has zero dipole. For time reversal, both curvature and the energy derivative are odd, leaving their product even. In contrast, the occupied-curvature integral defining intrinsic linear Hall conductivity remains odd. These statements require the full Hamiltonian and state to possess the symmetry; structural symmetry alone is insufficient when moments break it.

For the nonmagnetic equal-hopping control, a $120^\circ$ rotation $R_3$ would require $\vect D=R_3\vect D$. Since $R_3$ has no nonzero invariant vector in the plane, $\vect D=0$. Equivalently, the sum of a vector contribution and its two rotated partners vanishes. The $C_s$ control removes this rotational cancellation while keeping the mirror cancellation of $D_x$.

\subsection{Spin-polynomial representation}\label{S-sec:expansion}
Define the six-component input $\vect q=(s_{Ax},s_{Ay},s_{Az},s_{Bx},s_{By},s_{Bz})$. Under global spin reversal, the Hamiltonian is mapped by time reversal to the opposite momentum. Relabeling bands where necessary gives
\begin{equation}
 E_\nu(-\vect k;-\vect q)=E_\nu(\vect k;\vect q),\qquad
 \Omega_{\nu z}(-\vect k;-\vect q)=-\Omega_{\nu z}(\vect k;\vect q).
\end{equation}
The energy derivative in the Fermi-surface expression for $\vect D$ also changes sign. Hence
\begin{equation}
 \vect D(-\vect q)=\vect D(\vect q),\qquad
 \sigma_{xy}(-\vect q)=-\sigma_{xy}(\vect q).\label{S-eq:TR}
\end{equation}
The target parity requires even total degree in the spin expansion. The even parity distinguishes this response from the spin-reversal-odd ordinary anomalous Hall conductivity.

For a nonnegative integer multi-index $\vect p$, introduce the monomial
\begin{equation}
 I_{\vect p}(\vect q)=\prod_{r=1}^{6}q_r^{p_r},\qquad
 |\vect p|=\sum_rp_r\in\{0,2,4,6\}.\label{S-eq:monomials}
\end{equation}
Its mirror character is $(-1)^{p_1+p_3+p_4+p_6}$. The $x$ sector contains monomials with odd $p_1+p_3+p_4+p_6$, and the $y$ sector contains those with even $p_1+p_3+p_4+p_6$. The fitted representation is
\begin{equation}
 \widehat D_a(\vect q)/a=b_a+\sum_{\vect p\in\mathcal P_a}c_{a\vect p} I_{\vect p}(\vect q),\qquad a=x,y,\label{S-eq:polynomial}
\end{equation}
with $b_x=0$ and the sums restricted to nonconstant monomials. The constant is represented only once, through $b_y$, in the $y$ sector. Enumeration gives 280 raw $x$ features and 330 raw $y$ features, the latter including the constant. The family contains every monomial of the chosen degrees and characters, but it is not a claim that degree six represents the exact transport function.

The unit-length constraints $\vect s_A^2=\vect s_B^2=1$ produce dependencies between raw columns. For example, multiplying an allowed monomial by $s_{Ax}^2+s_{Ay}^2+s_{Az}^2$ yields the same function on the spin sphere. Consequently, different raw coefficients can represent identical predictions. PCA is used to remove numerical redundancy before supervised selection. Coefficients in the remaining basis describe a transport surrogate; they are not unique microscopic magnetic interactions.

Figure~\ref{S-fig:descriptor} connects the symmetry rule to an explicit orientation function, $I_x=s_{Ay}s_{Bx}$. Reversing both spins leaves it unchanged, whereas applying the mirror changes its sign. Panel (a) defines the two moment directions in their separate meridional planes, spanned by $\hat{\vect z}$ and $\boldsymbol\rho_i=\cos\phi_i\,\hat{\vect x}+\sin\phi_i\,\hat{\vect y}$. The visualization uses $\theta_A=\theta_++\theta_-$ and $\theta_B=\theta_+-\theta_-$ at fixed azimuths $\phi_A=0.37$ and $\phi_B=1.13$ rad.

In Fig.~\ref{S-fig:descriptor}(b), the red and blue regions correspond to the two signs of the product $s_{Ay}s_{Bx}$; the nodal lines occur when either factor vanishes. The white circle locates the spin pair drawn in panel (a), making the connection between moment orientation and a descriptor value explicit. The full mirror operation also changes the azimuths, so its character follows from the spin transformation, not from reflecting this fixed-azimuth image alone. This example explains how a globally reversal-even response can still vary strongly with relative and absolute moment orientation. The map is one basis function on a two-dimensional slice of the four-dimensional configuration space; its lobes are not regions of calculated Hall current. The band calculation and fitted coefficients determine how such allowed functions combine into the dipole studied in the main text.

\paperfigure[180mm]{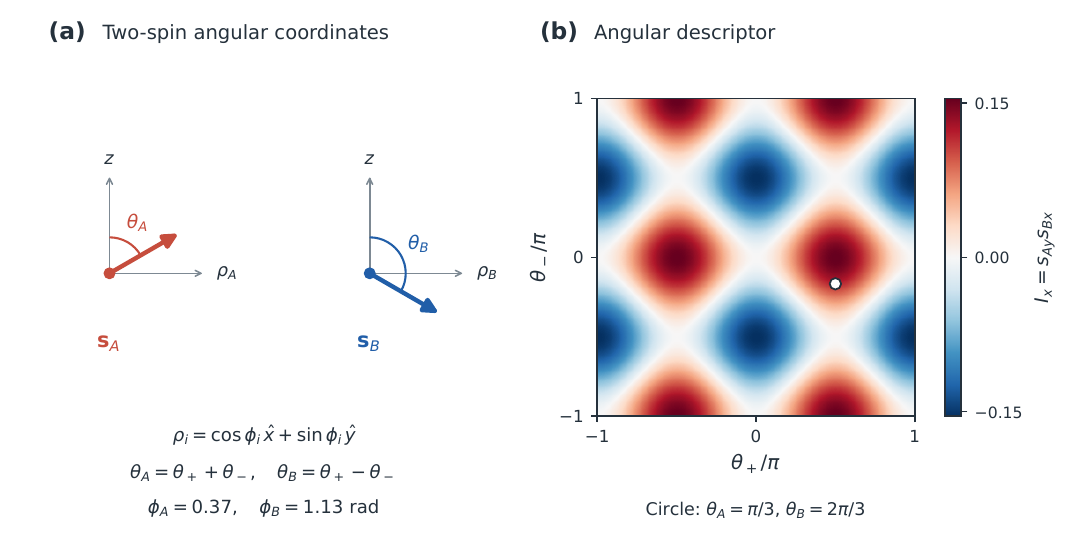}{Covariant descriptor $I_x=s_{Ay}s_{Bx}$. (a) Spin angles in the separate local $(\rho_i,z)$ planes at fixed azimuths $\phi_A=0.37$ and $\phi_B=1.13$ rad. (b) The descriptor on the $(\theta_+,\theta_-)$ slice, with angular axes in units of $\pi$ and a color bar showing its actual dimensionless value. The white circle corresponds to the arrows in (a), with $\theta_A=\pi/3$ and $\theta_B=2\pi/3$. This feature is even under simultaneous reversal of the spins and odd under the lattice mirror. The color represents a basis function, not a transport coefficient.}{descriptor}

\subsection{Site exchange and chirality}

The vector chirality $\vect\xi=\vect s_A\times\vect s_B$ is even under global spin reversal and odd under site exchange. It remains a useful coordinate for discussing noncollinearity. However, site exchange is not a symmetry of \eqref{S-eq:H} at fixed nonzero $\Delta$: it interchanges the two onsite energies. The response sectors are consequently fixed by the mirror characters of $D_x$ and $D_y$; site-exchange parity is not imposed as a symmetry of this Hamiltonian.

This distinction also limits a physical interpretation of the fitted features. An allowed bilinear can be written as a sum of exchange-even and exchange-odd parts; for instance,
\begin{equation}
 s_{Ay}s_{Bx}=\tfrac12(s_{Ay}s_{Bx}+s_{Ax}s_{By})-\tfrac12\xi_z.
\end{equation}
A coefficient multiplying this monomial does not isolate a chirality mechanism. The response may depend on spin orientation relative to the lattice, on correlations between the sites and on noncollinearity simultaneously. The symmetry construction includes these possibilities without assigning an unsupported microscopic interpretation to a particular fitted coefficient.

\subsection{Net and staggered magnetic coordinates}\label{S-sec:magneticcoordinates}
Define
\begin{equation}
 \vect M=\frac{\vect s_A+\vect s_B}{2},\qquad
 \vect L=\frac{\vect s_A-\vect s_B}{2},\qquad
 \vect M\cdot\vect L=0,\qquad M^2+L^2=1.\label{S-eq:ML}
\end{equation}
The last two identities follow from the unit lengths of both moments. Each coordinate is axial and odd under global spin reversal. Quadratic expressions in $M_iM_j$, $L_iL_j$, and $M_iL_j$ are therefore even under reversal. Their allowed response component is selected by the mirror character: for example, $M_xM_y$ is allowed in $D_x$, whereas $M_x^2$, $L_z^2$, and $M_xL_x$ are allowed in $D_y$. Site exchange would send $\vect L\mapsto-\vect L$ while preserving $\vect M$, but it is not a symmetry at fixed nonzero $\Delta$. It cannot be used to discard all mixed terms.

These are examples of permitted dependences, not a numerical identification of dominant contributions. The identities in Eq.~\eqref{S-eq:ML} also generate redundancies. A degree-six representation is an expansion in orientation functions on the constrained spin space, not a perturbation expansion in small $J$ or a proof of six-spin interactions. Similarly, $\vect s_A\times\vect s_B=-2\vect M\times\vect L$ is a useful coordinate but does not isolate the origin of a fitted bilinear.
\subsection{Bulk averaging over related magnetic domains}\label{S-sec:domains}
Consider macroscopic fractions $p$ and $1-p$ of two configurations related by global reversal. Assume identical Hamiltonian parameters, carrier conditions, relaxation time, and local electric field in the two domains, and retain only their bulk contributions. The time-reversal mapping makes their densities identical at a common chemical potential, so it also gives identical chemical potentials at common filling. Equation~\eqref{S-eq:TR} then gives
\begin{align}
 \overline{\vect D}&=p\vect D(\vect q)+(1-p)\vect D(-\vect q)=\vect D(\vect q),\\
 \overline{\sigma}_{xy}&=p\sigma_{xy}(\vect q)+(1-p)\sigma_{xy}(-\vect q)
 =(2p-1)\sigma_{xy}(\vect q).
\end{align}
The BCD currents add, while equal reversed populations cancel the intrinsic linear Hall current. This conclusion concerns the specified uniform-field bulk average, not a complete effective-medium conductivity or an open-circuit Hall voltage. Domain-wall scattering, spatially varying fields, contact geometry, and current redistribution have not been computed.

Mirror-related domains behave differently. Under the same averaging assumptions, Eqs.~\eqref{S-eq:Dmirror} give
\begin{equation}
 \overline{D}_x=(2p-1)D_x(\vect q),\qquad
 \overline{D}_y=D_y(\vect q).
\end{equation}
Equal mirror-domain fractions remove the component made available by mirror breaking while retaining the mirror-even component. 
\FloatBarrier
\section{Berry curvature, dipole, and occupation constraint}\label{S-sec:transport}
\subsection{Response definitions and normalization}\label{S-sec:responsedefinitions}

The definitions follow semiclassical Berry-curvature transport~\cite{SM-Sundaram1999,SM-Xiao2010,SM-Sodemann2015}. The conventions are the Berry connection $\vect{\mathcal A}_\nu=i\langle u_\nu|\nabla_{\vect k}u_\nu\rangle$, electron charge $-e$ with $e>0$, and $\epsilon_{xyz}=+1$. For a nondegenerate band, the curvature is evaluated as
\begin{equation}
 \Omega_{\nu z}(\vect k)=-2\operatorname{Im}\sum_{\lambda\ne\nu}
 \frac{\langle u_\nu|\partial_{k_x}H|u_\lambda\rangle
 \langle u_\lambda|\partial_{k_y}H|u_\nu\rangle}
 {(E_\nu-E_\lambda)^2}.\label{S-eq:berry}
\end{equation}
 Individual-band curvature is used away from exact degeneracies; the integration checks concern the band-summed transport response.

With $f_\nu=[1+\exp((E_\nu-\mu)/\kbT)]^{-1}$ and $[\dd k]=\dd^2k/(2\pi)^2$, the dipole is
\begin{align}
 D_a&=\sum_\nu\int_{\BZ}[\dd k]\,f_\nu\,\partial_{k_a}\Omega_{\nu z}\label{S-eq:sea}\\
 &=\sum_\nu\int_{\BZ}[\dd k]\,\Omega_{\nu z}\,
       \partial_{k_a}E_\nu\left(-\partial_{E_\nu}f_\nu\right).\label{S-eq:surface}
\end{align}
Integration by parts on the reciprocal torus gives the second expression. It emphasizes a thermally broadened Fermi-surface contribution, since $-\partial_E f=f(1-f)/\kbT$. The computed dipole has dimensions of length and is reported as $D_a/a$. The electronic temperature is approximately $580$ K; it regularizes the occupation profile, not the interband curvature denominator.

Let $\av{g}=\int_0^1\dd u\int_0^1\dd v\,g(u\vect b_1+v\vect b_2)$, where $\vect a_i\cdot\vect b_j=2\pi\delta_{ij}$. Then
\begin{equation}
 \int_{\BZ}[\dd k]g=\frac{\av{g}}{A_{\rm cell}},\quad
 n=\frac12\av{\sum_\nu f_\nu},\quad
 \frac{\sigma_{xy}}{e^2/h}=-\frac{2\pi}{A_{\rm cell}}\av{\sum_\nu f_\nu\Omega_{\nu z}}.\label{S-eq:normalization}
\end{equation}
All four bands are included once. The filling $n$ is electrons per site, with capacity $0\le n\le2$. For the fixed-filling calculation, $\mu$ is solved separately for each spin pair so that $n=n_0$.

\subsection{Fermi contours and the energy-resolved curvature weight}\label{S-sec:spectral}
For regular constant-energy contours, define
\begin{align}
 W_a(\epsilon;\vect q)
 &=\sum_\nu\int_{\BZ}[\dd k]\,
 \Omega_{\nu z}\,\partial_{k_a}E_\nu\,
 \delta(\epsilon-E_\nu)\label{S-eq:spectralweight}\\
 &=\sum_\nu\oint_{E_\nu=\epsilon}\frac{\dd\ell}{(2\pi)^2}
 \Omega_{\nu z}\frac{\partial_{k_a}E_\nu}{|\nabla_{\vect k}E_\nu|}.\label{S-eq:contour}
\end{align}
The second line is the coarea formula. The ratio of derivatives is the component of the contour normal toward increasing energy. Summing all contours is essential: pocket contributions can reinforce or cancel, and a large curvature on one band does not determine the sign of the full dipole. At a critical contour where the gradient vanishes, the first line defines the appropriate distributional expression; the regular-contour form must be interpreted as a limit.

The finite-temperature response is
\begin{equation}
 D_a(\mu,T;\vect q)=\int_{-\infty}^{\infty}\dd\epsilon\,
 W_a(\epsilon;\vect q)
 \frac{1}{4\kbT}\operatorname{sech}^2\!\left(\frac{\epsilon-\mu}{2\kbT}\right).\label{S-eq:convolution}
\end{equation}
The kernel integrates to one and tends to $\delta(\epsilon-\mu)$ as $T\to0$, giving the Fermi-contour expression in the main text. The finite-temperature calculation thus averages signed contributions across an energy window of order a few $\kbT$. At zero temperature, a completely filled isolated band has no BCD contribution, even if its integrated curvature is nonzero. The dipole is consequently distinct from a Chern number and is not quantized.

Small interband separations can enhance Eq.~\eqref{S-eq:berry}, but the dipole also depends on the matrix elements, contour orientation, and the weighting in Eq.~\eqref{S-eq:convolution}. Band-geometry studies identify these sensitivities in other systems~\cite{SM-Du2018,SM-Facio2018,SM-Battilomo2019}. Equations~\eqref{S-eq:spectralweight}--\eqref{S-eq:convolution} provide a physical interpretation of the reported calculations.

\subsection{Chemical-potential feedback at fixed filling}\label{S-sec:feedback}
Let $\lambda$ parameterize a smooth path of the two prescribed moments at fixed Hamiltonian parameters and electronic temperature. Write the occupation as $n(\mu,\lambda)$ and require $n(\mu(\lambda),\lambda)=n_0$. Differentiation gives
\begin{equation}
 \left(\frac{\dd\mu}{\dd\lambda}\right)_n
 =-\frac{(\partial_\lambda n)_\mu}{(\partial_\mu n)_\lambda}.\label{S-eq:mufeedback}
\end{equation}
For the noninteracting band occupations and the normalization in Eq.~\eqref{S-eq:normalization},
\begin{align}
 (\partial_\mu n)_\lambda
 &=\frac12\av{\sum_\nu(-\partial_{E_\nu}f_\nu)},\\
 (\partial_\lambda n)_\mu
 &=\frac12\av{\sum_\nu(\partial_{E_\nu}f_\nu)\partial_\lambda E_\nu}.
\end{align}
The first quantity is positive at the finite temperature considered and measures the change in filling with chemical potential. Applying the chain rule to the dipole yields
\begin{equation}
 \left(\frac{\dd D_a}{\dd\lambda}\right)_n
 =\left(\frac{\partial D_a}{\partial\lambda}\right)_\mu
 -\left(\frac{\partial D_a}{\partial\mu}\right)_\lambda
 \frac{(\partial_\lambda n)_\mu}{(\partial_\mu n)_\lambda}.\label{S-eq:Dfeedback}
\end{equation}
At fixed chemical potential, both curvature and band velocities can change with orientation. Fixed filling adds the motion of the occupation window needed to compensate the energy shifts. The feedback can reinforce or oppose the direct variation and has no symmetry-imposed sign. This explains why fixed filling need not simplify the orientation dependence. Near an incompressible zero-temperature regime, Eq.~\eqref{S-eq:mufeedback} requires care because a unique differentiable chemical potential may not be defined. The reported comparison is at finite temperature and at the stated metallic filling.

The numerical fixed-filling dataset solves the occupation equation for each configuration. The derivative terms in Eq.~\eqref{S-eq:Dfeedback} were not independently evaluated or fitted. The equation is an analytical interpretation of the distinction between the two numerical targets; their quantitative comparison is in Sec.~\ref{S-sec:fits}.

\FloatBarrier
\section{Nonlinear-current conventions and frequency dependence}\label{S-sec:current}
\subsection{Occupation shift, anomalous velocity, and the static current}\label{S-sec:boltzmann}
The relaxation-time Boltzmann equation for a spatially uniform occupation is
\begin{equation}
 \partial_t f_\nu-\frac{e}{\hbar}\vect E\cdot\nabla_{\vect k}f_\nu
 =-\frac{f_\nu-f_\nu^{(0)}}{\tau}.
\end{equation}
For a static weak field, its first-order solution and the anomalous velocity are
\begin{equation}
 f_\nu^{(1)}=\frac{e\tau}{\hbar}\vect E\cdot\nabla_{\vect k}f_\nu^{(0)},\qquad
 \vect v_\nu^{\rm an}=\frac{e}{\hbar}\vect E\times\hat{\vect z}\,\Omega_{\nu z}.
\end{equation}
Their contribution to current is $-e\sum_\nu\int[\dd k]\,\vect v_\nu^{\rm an}f_\nu^{(1)}$. Since $\nabla_{\vect k}f=(\partial_Ef)\nabla_{\vect k}E$, the negative occupation derivative combines with the negative electronic charge to give the sign below. This is the BCD term in the semiclassical expansion~\cite{SM-Sundaram1999,SM-Sodemann2015}; it does not include every possible second-order distribution or velocity contribution.

The resulting static sheet current is
\begin{equation}
 \vect j^{\BCD}_{\rm static}=\frac{e^3\tau}{\hbar^2}
       (\vect E\times\hat{\vect z})(\vect D\cdot\vect E).\label{S-eq:staticcurrent}
\end{equation}
Its sign follows directly from $\hbar\dot{\vect k}=-e\vect E$, anomalous velocity $(e/\hbar)\vect E\times\vect\Omega$, and the first-order occupation correction $(e\tau/\hbar)\vect E\cdot\nabla_{\vect k}f$. In particular, a field along $y$ gives $j_x=e^3\tau D_yE_y^2/\hbar^2$. This explicit convention avoids inferring a current sign from a curvature plot with an unspecified Berry-connection convention.

\subsection{Harmonics, frequency dependence, and the meaning of transversality}\label{S-sec:frequency}
For $\vect E(t)=\operatorname{Re}[\vect{\mathcal E}e^{i\omega t}]$, define $j_a^{(0)}=\operatorname{Re}[\chi_{abc}\mathcal E_b\mathcal E_c^*]$ and $j_a^{(2\omega)}(t)=\operatorname{Re}[\chi_{abc}\mathcal E_b\mathcal E_c e^{2i\omega t}]$. An unsymmetrized tensor whose contraction gives the BCD response is
\begin{equation}
 \chi_{abc}(\omega)=-\epsilon_{azc}
 \frac{e^3\tau}{2\hbar^2(1+i\omega\tau)}D_b.\label{S-eq:chi}
\end{equation}
Thus $\chi_{xyy}=+CD_y$ and $\chi_{yxx}=-CD_x$, where $C=e^3\tau/[2\hbar^2(1+i\omega\tau)]$. For a slow real cosine field, the rectified and second-harmonic amplitudes are each half the static response at the same peak field. Equations~\eqref{S-eq:staticcurrent} and \eqref{S-eq:chi} describe a weak-field intraband contribution. 

For a real linearly polarized field, the rectified amplitude contains the real factor $[1+(\omega\tau)^2]^{-1}$, while the second-harmonic amplitude has magnitude proportional to $[1+(\omega\tau)^2]^{-1/2}$ and a phase lag from $(1+i\omega\tau)^{-1}$.

The static expression satisfies $\vect j^{\BCD}\cdot\vect E=0$. It coexists with longitudinal current and scattering, so the identity is not a statement that the whole sample is dissipationless or that energy conversion has unit efficiency. 

For an arbitrary reference field $E_0$, define $j_0=e^3\tau aE_0^2/\hbar^2$. The nonmagnetic control has $D_x=0$ and $D_y/a=-0.0127421406345$ at $\mu_0$. Consequently, $j_x/j_0=(D_y/a)(E_y/E_0)^2$ for a field along $y$. For a rotating static field $\vect E=E_0(\cos\theta,\sin\theta)$, the normalized components are $(D_y/a)\sin^2\theta$ and $-(D_y/a)\sin\theta\cos\theta$. These equations give the field and angular panels in the main text.

For a slow cosine drive along $y$, the dc and second-harmonic amplitudes are each $D_yj_0/(2a)$. The frequency-dependent magnitude of the kinetic factor is $[1+(\omega\tau)^2]^{-1/2}$, while its real part is $[1+(\omega\tau)^2]^{-1}$. 
\FloatBarrier
\section{Brillouin-zone integration and numerical validation}\label{S-sec:numerics}
The final labels are computed by adaptive quadrature. A $16\times16$ partition of the fractional reciprocal cell is refined using a comparison between tensor-product $3\times3$ and $5\times5$ Gauss rules, supplemented by a small-gap trigger. The stopping tolerances are $10^{-6}$ for $D_a/a$, $10^{-8}$ for filling and $5\times10^{-6}$ for $\sigma_{xy}/(e^2/h)$. The refinement rule and independent checks are detailed below. Adaptive refinement is an established tool for sharply structured Berry-curvature integrals~\cite{SM-Tsirkin2021}.

\subsection{Adaptive refinement and numerical degeneracy guard}

The adaptive integrator operates on $[0,1)^2$ in fractional reciprocal coordinates. Each active rectangular cell has independent $3\times3$ and $5\times5$ Gauss estimates for both dipole components at all requested chemical potentials, the filling and the linear Hall conductivity. The fine estimate is used for the integral, while the absolute difference supplies a local empirical error indicator. Cell errors are summed for each observable. Refinement continues until every summed indicator is below its specified absolute tolerance and no active cell is flagged by the small-gap condition.

The gap indicator compares the smallest adjacent-band separation sampled in a cell with a scale proportional to the cell's reciprocal-space width. In the implementation, energies are in eV and lengths in units of $a$, and the proportionality coefficient is $1.2\eV\,a$. In these units the trigger is
\begin{equation}
 g_{\min}<1.2\,\max(\Delta u,\Delta v)\,
        \max( |\vect b_1|,|\vect b_2| ).\label{S-eq:gaptrigger}
\end{equation}
This is a refinement heuristic, not a rigorous estimate of a local band velocity or an integral-error theorem. Band pairs whose mean energy is farther than $30\kbT$ from every chemical potential in the call are omitted only from this forced-refinement trigger. Their contributions remain in all integrands and error estimates. Selected cells are divided into four children. The implementation limits depth to 16 subdivisions beyond the initial grid and total evaluations to three million points per call; the reported final calculations meet the stopping criteria rather than terminating at these limits.

In the curvature sum, diagonal terms are excluded and energy differences below $10^{-10}\eV$ are masked to avoid division at numerical degeneracy. This numerical guard is not a lifetime broadening and does not replace the gap denominator by a Lorentzian. Individual-band curvature at exact degeneracies is not assigned a physical value by this procedure. Shifted uniform meshes and interior Gauss nodes avoid the principal symmetry-degeneracy points of the nonmagnetic controls. Applications involving extended degeneracies would require a corresponding degenerate-subspace treatment and new validation.

Fixed-filling calculations solve the occupation constraint on the initial uniform mesh and then check it with the adaptive rule. The resulting chemical potentials pass this check without requiring an update in the reported run. The quoted $6.95\times10^{-14}$ discrepancy is the final numerical occupation residual, while the density quadrature has its separate error indicator. For the independent Fermi-sea check, both Cartesian curvature derivatives are approximated by central differences of freshly diagonalized Hamiltonians at $\vect k\pm10^{-6}a^{-1}\hat{\vect e}_a$.

\subsection{Convergence audit and independent checks}

The initial $192\times192$ shifted uniform grid is inadequate for a small subset of configurations. A 40-configuration audit combines 24 deterministic coverage samples with eight small-gap and eight large-response stress cases. Comparisons of $96$, $192$, $320$ and $480$ points per reciprocal direction show nonmonotonic changes near narrow avoided crossings. 

Sample ID~747 provides a clear stress case. At $\mu=0.90\eV$, the $192\times192$ mesh gives $D_x/a=-0.108335$, whereas adaptive quadrature gives $+0.00522829$. The $320\times320$ and $480\times480$ values are $+0.0138886$ and $-0.0242723$, respectively. An inspected local avoided crossing has a gap of about $1.06\times10^{-3}\eV$ near energy $1.092\eV$. Even though it lies above the chemical potential, its curvature can affect the finite-temperature integral. This example is evidence of under-resolved quadrature, not a newly identified phase or a discontinuous physical transition.

Every one of the 1024 configurations was therefore recomputed with the adaptive method; no outlier was deleted and no target was clipped. All met the stated stopping criteria. The final run used 209,999,504 band-evaluation points in total, with 169,048--242,216 points per configuration. 

The 40 audited configurations were then recomputed with a dipole tolerance of $2\times10^{-8}$, fifty times tighter than the production setting. The maximum observed change at $\mu_0$ was $2.42\times10^{-12}a$, and the largest change over all eleven energies was $1.02\times10^{-10}a$. The fixed-filling dipoles changed by at most $2.27\times10^{-12}a$. 

An independent evaluation of \eqref{S-eq:sea}, using central finite differences of the curvature with $\delta k=10^{-6}a^{-1}$ and adaptive integration, agrees with \eqref{S-eq:surface} within $5.60\times10^{-12}a$ at $\mu_0$ for IDs~747 and 0. Analytic Hamiltonian derivatives agree with finite differences within $2.42\times10^{-10}\eV\,a$ in the derivative audit. Direct covariance tests on six configurations give maximum differences of $1.16\times10^{-15}a$ under global spin reversal and $1.32\times10^{-15}a$ under the mirror; the linear Hall sign-reversal difference is $6.25\times10^{-17}e^2/h$.

\paperfigure[180mm]{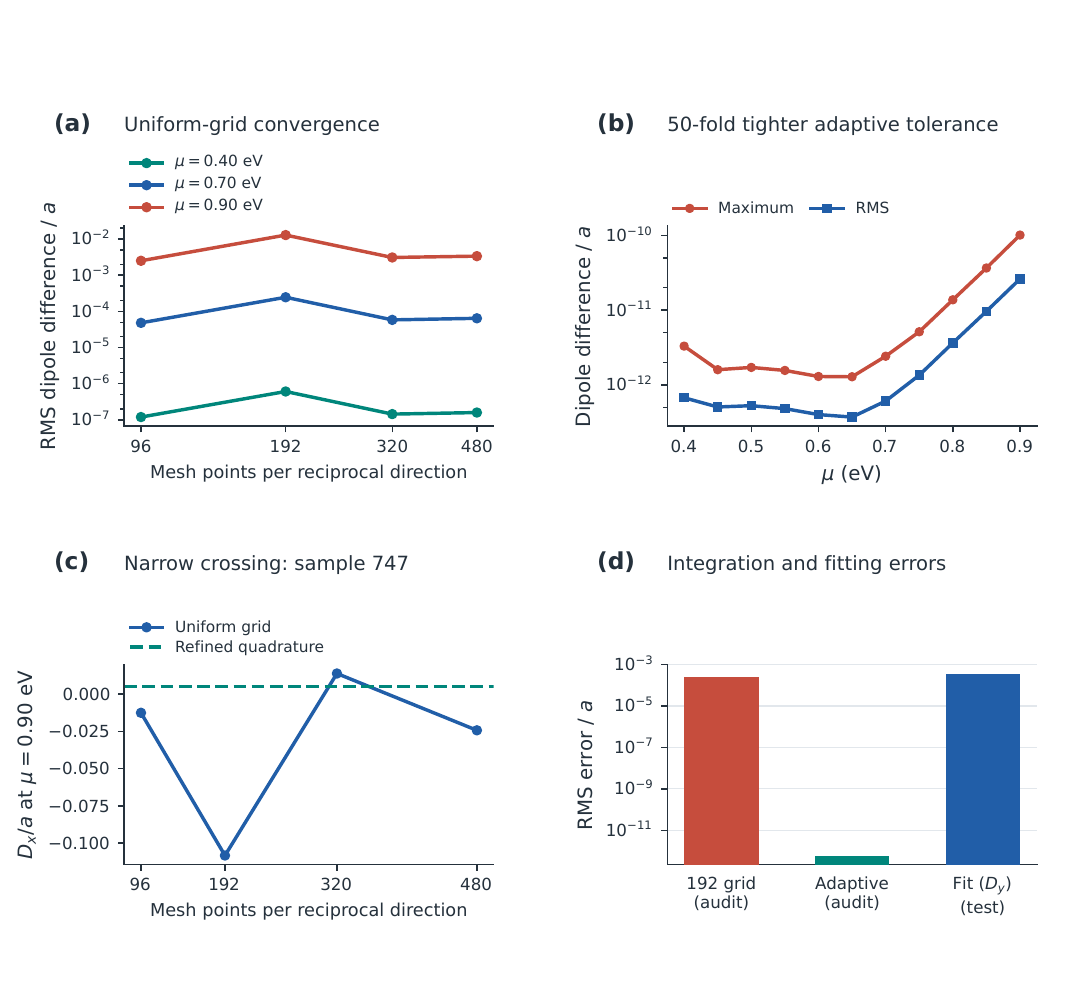}{Numerical validation. (a) Uniform-grid integration compared with refined quadrature. (b) Sensitivity to a fifty-fold tighter adaptive tolerance. (c) The ID~747 stress case. (d) Separate scales of quadrature and regression error. The final learned targets use adaptive quadrature. The 40-configuration audit deliberately includes difficult cases and is not a random-confidence bound for every possible spin configuration.}{convergence}

Figure~\ref{S-fig:convergence}(a) shows why a larger uniform grid alone is insufficient: its error relative to refined quadrature is nonmonotonic, especially at the highest chemical potential. Panel (c) gives the physical consequence in the ID~747 stress case, where coarse integration even changes the sign of $D_x$. Such a sign would incorrectly reverse the inferred quadratic current. The small changes under the tighter adaptive tolerance in Fig.~\ref{S-fig:convergence}(b) instead establish stability of the refined labels over the audited energy range.

Figure~\ref{S-fig:convergence}(d) places the initial uniform-grid discrepancies, adaptive stability differences, and final fitting errors on a common logarithmic scale. The fitting errors are approximately $10^{-4}a$ at $\mu_0$, several orders of magnitude above the observed adaptive changes. The bars compare distinct diagnostics and sample sets, rather than a pointwise decomposition of one error. Together they support interpreting the remaining fit residuals as limitations of the chosen surrogate at the audited conditions, rather than as the unresolved uniform-grid artifacts of the preliminary labels. This is the numerical basis for assigning physical meaning to the orientation dependence in main-text Fig.~\ref{fig:predictions}. 

\begin{table}[!htbp]
\caption{Selected numerical checks. Values are maximum observed absolute differences over the stated audit, not guaranteed bounds outside it.}\label{S-tab:validation}
\centering\small
\begin{tabular}{p{84mm}r}\toprule
Check&Maximum difference\\\midrule
Adaptive tolerances $10^{-6}$ vs $2\times10^{-8}$, $\mu_0$, 40 IDs&$2.42\times10^{-12}a$\\
Same audit over all eleven chemical potentials&$1.02\times10^{-10}a$\\
Same audit at fixed filling&$2.27\times10^{-12}a$\\
Surface vs finite-difference sea dipole, IDs 747 and 0&$5.60\times10^{-12}a$\\
Spin-reversal covariance of direct dipoles, six IDs&$1.16\times10^{-15}a$\\
Mirror covariance of direct dipoles, six IDs&$1.32\times10^{-15}a$\\
Linear Hall sign under spin reversal, six IDs&$6.25\times10^{-17}e^2/h$\\
Hamiltonian derivative vs finite difference&$2.42\times10^{-10}\eV\,a$\\\bottomrule
\end{tabular}
\end{table}

\subsection{Direct-control grids and display conventions}
The nonmagnetic $C_s$ and $C_{3v}$ control curves, and the magnetic sample ID 0 and its reversed partner, use a $480\times480$ shifted reciprocal mesh. The electronic-structure maps use a $192\times192$ mesh. The plotted band path is a geometric $\Gamma$--$M$--$K$--$\Gamma$ path on the honeycomb reciprocal lattice; it does not imply that the anisotropic Hamiltonian retains threefold symmetry. In the signed maps, only the displayed color range saturates at the largest 0.5\% of absolute values. The final 1024-configuration regression dataset uses the adaptive labels described above.

\FloatBarrier
\section{Spin sampling and the held-out configuration set}\label{S-sec:sampling}

For each site, $z$ and $\phi$ are sampled uniformly in $[-1,1]$ and $[0,2\pi)$, respectively. The corresponding unit vector is
\begin{equation}
 \vect s=\left(\sqrt{1-z^2}\cos\phi,\sqrt{1-z^2}\sin\phi,z\right).
\end{equation}
The two sites are sampled independently, giving uniform surface-area measure on $S^2\times S^2$. The 1024 spin pairs are static input configurations. They are not obtained from a spin Hamiltonian, a thermal Boltzmann weight or a dynamical trajectory. In particular, the electronic temperature in table~\ref{S-tab:parameters} does not determine the probability of a spin orientation.

A random permutation divides the original configuration IDs into 768 training and 256 test groups before any symmetry augmentation. Sampling and splitting use seeds 20261017 and 31704, respectively. In the $x$ sector the mirror generates distinct rows with opposite feature and target signs; both partners remain in the same training group. Global spin reversal duplicates the even-degree feature row and target, and the $y$ mirror does likewise. Such identical copies are omitted. Test configurations remain excluded from preprocessing, hyperparameter selection and coefficient estimation.

Figures~\ref{S-fig:samples}(a) and \ref{S-fig:samples}(b) display the $A$ moments in the $xy$ projection and the $B$ moments in the $xz$ projection. Both include orientations throughout the projected disk, consistent with exploring rotations relative to all crystal axes rather than one prescribed rotation path. The greater apparent density near a rim is a consequence of projecting a uniform sphere, not evidence of a preferred magnetic orientation. The independently generated pairs span collinear and noncollinear neighborhoods without defining a thermodynamic magnetic phase. This sampling gives the held-out test in main-text Fig.~\ref{fig:predictions} its meaning as a test across moment orientations within the same Hamiltonian. At $\mu_0=0.70\eV$, the complete sample has $D_x/a$ between $-0.06586$ and $0.06876$, and $D_y/a$ between $-0.05888$ and $0.01949$. These are extrema of the finite sample, not rigorous extrema over all spin pairs.

\paperfigure[180mm]{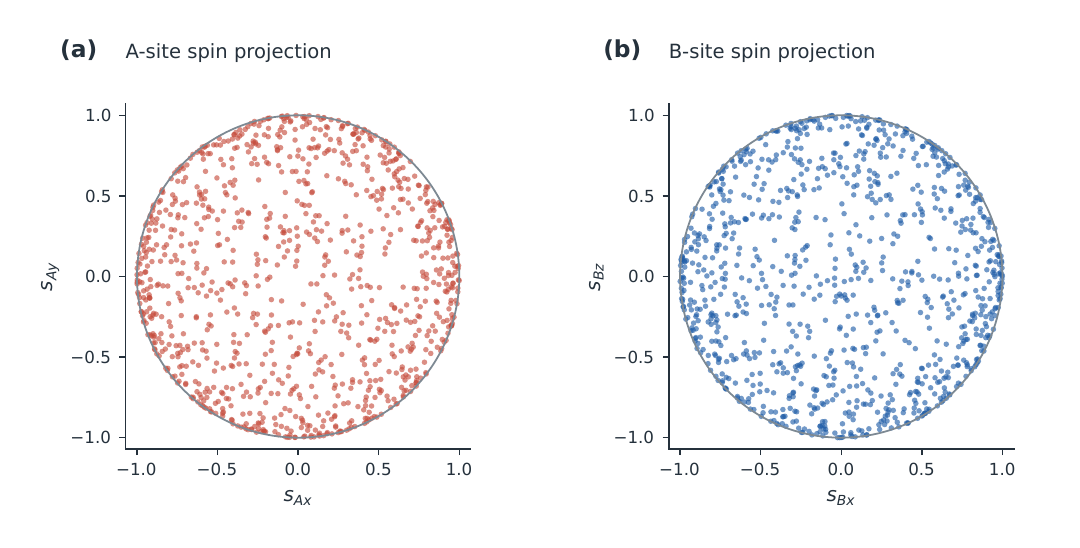}{Projections of the 1024 calculated spin pairs: (a) the $A$ moments in the $xy$ plane and (b) the $B$ moments in the $xz$ plane. Each spin was sampled independently with uniform azimuth and uniform $z$ component. The distribution is uniform on each sphere, not on its planar projection. The points are input configurations and carry no assumed equilibrium statistical weight.}{samples}

\FloatBarrier
\section{Regression pipeline and model selection}\label{S-sec:regression}

Figure~\ref{S-fig:workflow} separates the physical information entering the orientation map from the statistical steps used to represent it. The first box supplies dipoles from the Bloch eigenstates; symmetry then restricts which angular functions can describe each component, while the fitted coefficients determine their quantitative combination. The left branch chooses settings using training configurations, and the right branch reserves original configuration IDs until the final prediction test. Splitting before symmetry augmentation prevents a transformed copy of a test configuration from entering the fit. Thus the final agreement in main-text Fig.~\ref{fig:predictions} tests prediction for untrained orientations, rather than recovery of known symmetry partners. 

Each component is fitted separately. The feature matrix is first restricted to columns with nonzero training variance, standardized, and decomposed by singular-value decomposition. If the singular values are $s_j$, the explained-variance fraction is $r_j=s_j^2/\sum_\ell s_\ell^2$. Components with $r_j\le10^{-12}$ are discarded. This numerical-rank step prevents redundant spin monomials from generating spurious near-null PCs.

\paperfigure[180mm]{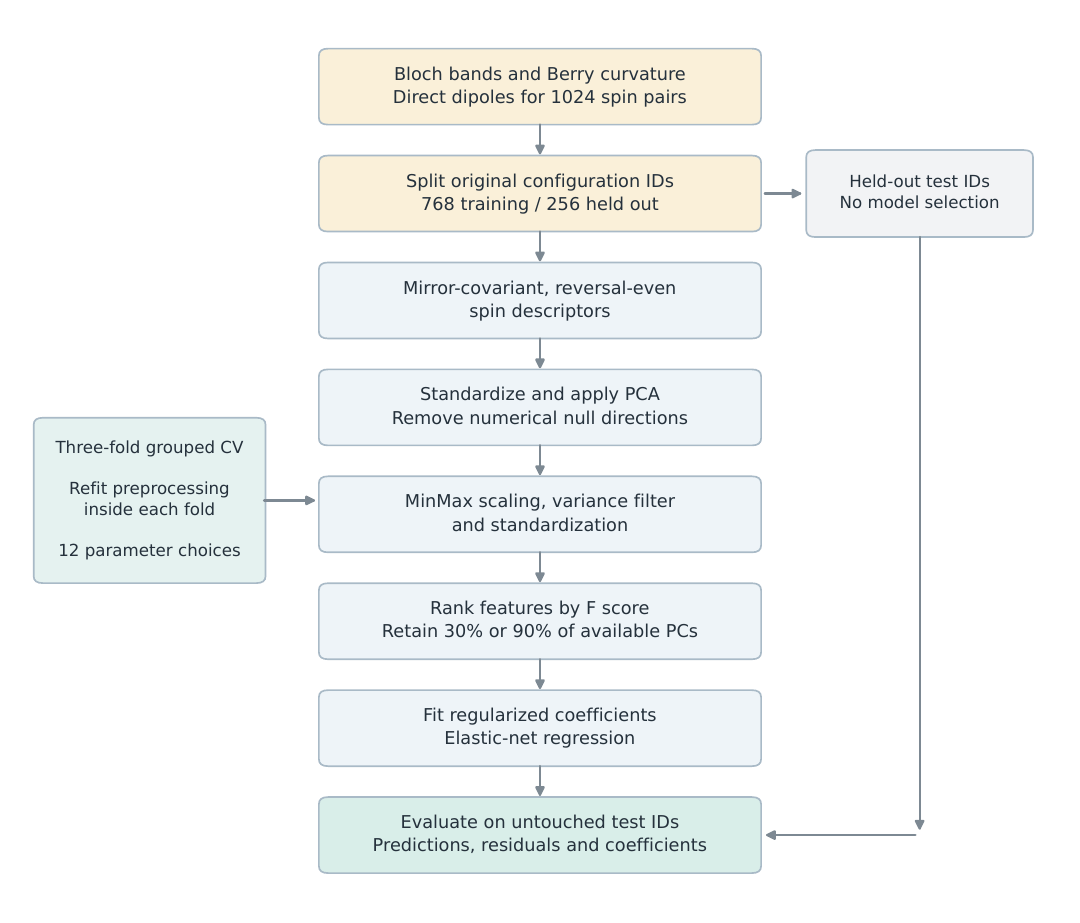}{Transport and regression workflow. Original configuration IDs are split before symmetry augmentation. Preprocessing, PCA and supervised feature ranking are fitted within the training data, including each of the three cross-validation folds. The 256 test configurations are used only for the final evaluation. The two feature budgets retain 30\% and 90\% of the available nonconstant PCs.}{workflow}

The surviving PC scores are mapped by training-set MinMax scaling, filtered at variance $10^{-5}$, and standardized again. For each target, univariate F-regression scores rank the remaining components. A feature budget $f$ retains $\lceil fR\rceil$ components, where $R$ is the retained rank and $f\in\{0.3,0.9\}$. The F statistic measures an individual linear association with the target; it is neither a probability nor a unique measure of a component's physical importance.

The selected features are fitted using squared-error stochastic gradient descent with elastic-net regularization~\cite{SM-Zou2005,SM-Pedregosa2011}. In standardized target units, the objective has the form
\begin{equation}
 \mathcal L(\vect\beta)=\frac{1}{2N_{\rm aug}}\sum_i
 (\widetilde y_i-\widetilde{\vect Z}_i\cdot\vect\beta)^2
 +\lambda\left[\rho\|\vect\beta\|_1+\frac{1-\rho}{2}\|\vect\beta\|_2^2\right].\label{S-eq:loss}
\end{equation}
Target scaling uses training data only. For $D_y$, the training mean is removed and restored, giving the allowed affine offset. For $D_x$, transformed features are referenced to the image of the zero raw-feature vector, which ensures an exactly zero intercept in the back-transformed physical polynomial. 

Three-fold cross-validation on original training IDs chooses $\lambda\in\{10^{-5},10^{-4},10^{-3}\}$, $\rho\in\{0.15,0.85\}$ and initial learning rate $\eta_0\in\{0.003,0.01\}$. All preprocessing is refitted within each fold, and symmetry partners remain together. The learning-rate schedule is inverse scaling with exponent $0.25$, the maximum iteration count is 6000, the optimization tolerance is $10^{-8}$, and the no-improvement patience is 35 iterations. No convergence warning was emitted for the final models. The selected parameters are recorded in Sec.~\ref{S-sec:reproducibility}.

For the energy sweep, the training PCA basis and hyperparameters selected at $\mu_0$ are reused. F ranking and regression coefficients are recalculated from training labels at each energy. Thus each energy has a separately trained spin-to-dipole map; the study does not test prediction at an untrained chemical potential. 

Prediction quality is quantified by held-out $R^2$, root-mean-square error (RMSE) and maximum absolute error. For test targets $y_i$, predictions $\widehat y_i$ and test mean $\bar y$,
\begin{equation}
 R^2=1-\frac{\sum_i(y_i-\widehat y_i)^2}{\sum_i(y_i-\bar y)^2},\qquad
 \mathrm{RMSE}=\sqrt{\frac1{N_{\rm test}}\sum_i(y_i-\widehat y_i)^2}.\label{S-eq:metrics}
\end{equation}
Residuals in the figures are $D_a-\widehat D_a$. A high $R^2$ describes variance explained on this test set.

\FloatBarrier
\section{Raw features, principal components, and statistical diagnostics}\label{S-sec:features}
Figure~\ref{S-fig:xfeatures} displays the first 36 canonical monomials for $D_x$ on the angular slice defined in Sec.~\ref{S-sec:expansion}. The alternating signs and nodal lines illustrate the angular structure available to represent a response activated by mirror-breaking moments. Every function is odd under the full structural mirror and even under simultaneous spin reversal; no constant is present. Consequently, the representation preserves the exact zero of $D_x$ for a mirror-invariant configuration even when many angular terms are retained.

Figure~\ref{S-fig:yfeatures} shows the contrasting $D_y$ sector. Its first panel is the uniform degree-zero function, followed by orientation-dependent terms with even mirror character. The constant is allowed because a mirror-preserving state can retain a dipole normal to the mirror, as the nonmagnetic control in main-text Fig.~\ref{fig:controls} demonstrates. The constant of a fitted finite-$J$ polynomial, however, should not be identified with the separately calculated $J=0$ response. Removing exchange changes the Hamiltonian, whereas the learned offset describes the sampled orientation dependence at fixed exchange. Similar-looking patterns in these two fixed-azimuth slices do not make the full functions equivalent: their symmetry characters concern transformations of all spin coordinates.

\paperfigure[180mm]{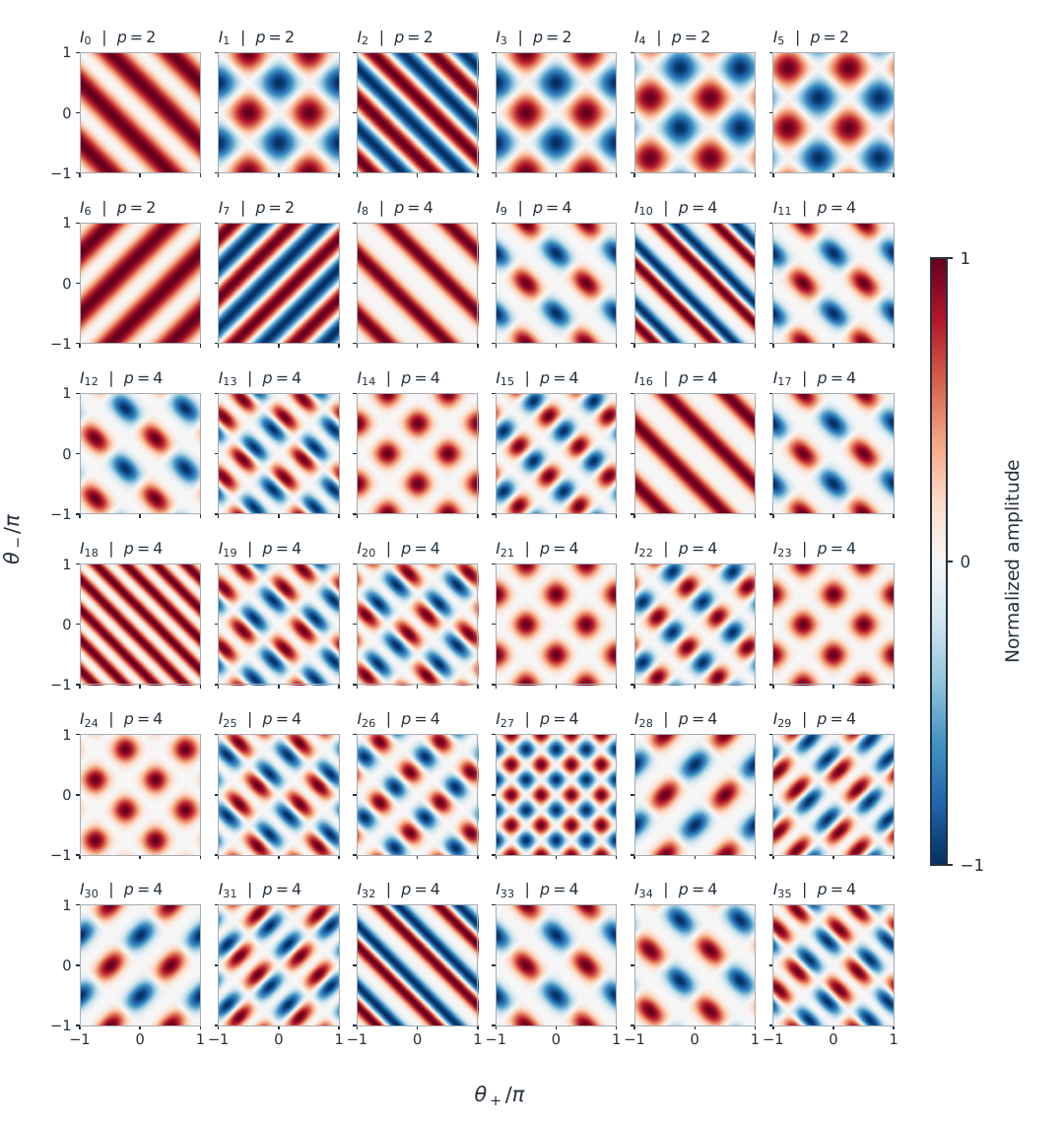}{First 36 canonical features for $D_x$, ordered by increasing even total degree $p$ and a fixed monomial enumeration. All are odd under the structural mirror and even under global spin reversal. Each angular map is divided by its own maximum absolute value. The axes show $\theta_+/\pi$ and $\theta_-/\pi$ at fixed azimuths $\phi_A=0.37$ and $\phi_B=1.13$ rad. These are component covariants, not an exchange-symmetry sector.}{xfeatures}

\paperfigure[180mm]{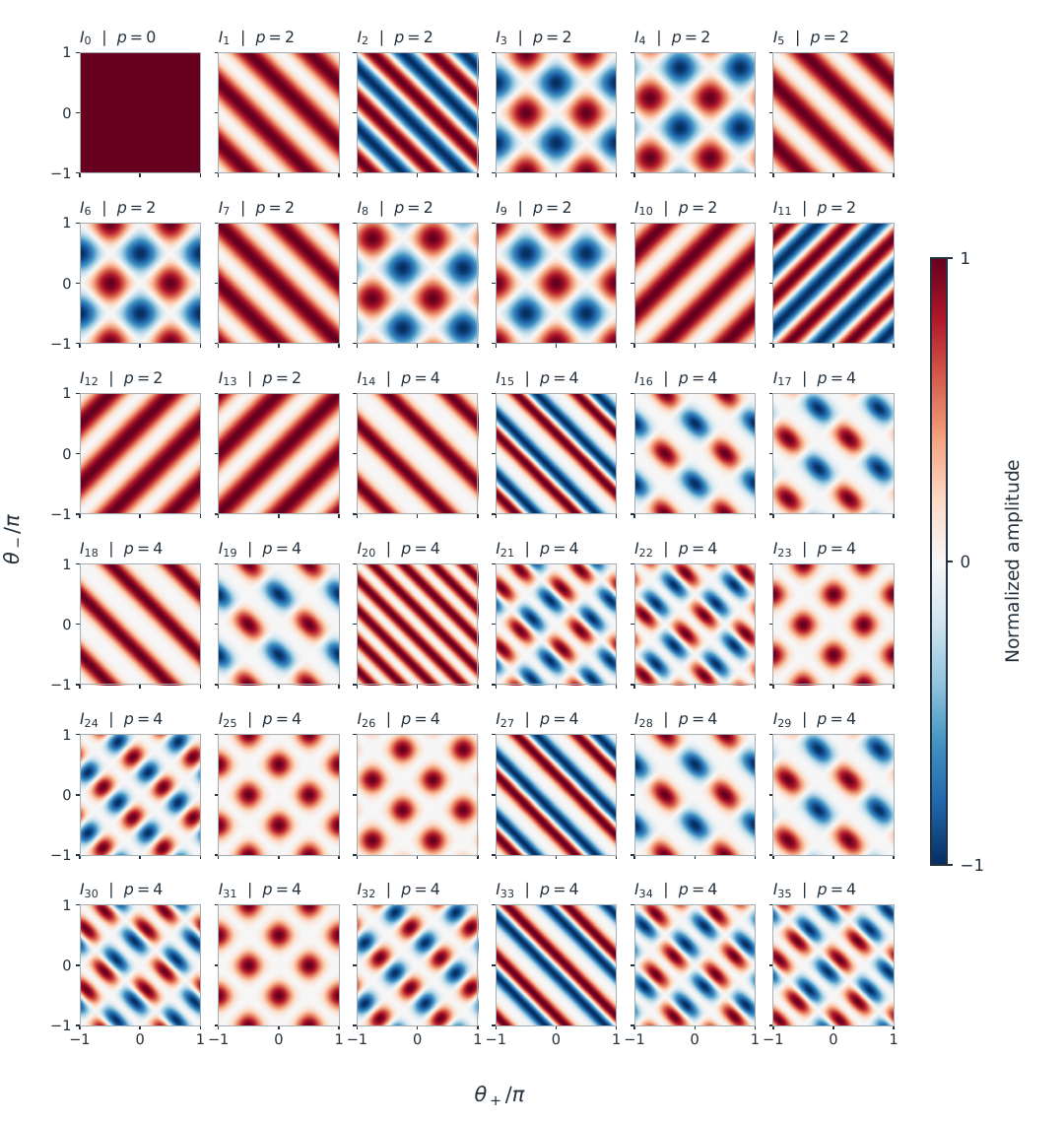}{First 36 canonical features for $D_y$, including the allowed constant. All are even under the mirror and global spin reversal. The angular slice and independent panel normalization are the same as in figure~\ref{S-fig:xfeatures}.}{yfeatures}

\subsection{Principal-component transformation}\label{S-sec:pcaresults}
The raw monomials have substantial statistical correlations as well as exact dependencies on the unit spheres. After centering and removal of null directions, the numerical ranks are 160 for $D_x$ and 175 for the nonconstant $D_y$ sector. Table~\ref{S-tab:features} records the reduction. The apparent difference between 330 raw $y$ columns and 175 retained PCs includes removal of the constant column; its contribution is restored through the allowed affine offset.

\begin{table}[!htbp]
\caption{Feature counts. Budgets refer to the retained nonconstant PCs and are rounded upwards. The raw $D_y$ count includes one constant.}\label{S-tab:features}
\centering\small
\begin{tabular}{lrrrr}\toprule
Component&Raw columns&Retained PCs&30\% budget&90\% budget\\\midrule
$D_x$&280&160&48&144\\
$D_y$&330&175&53&158\\\bottomrule
\end{tabular}
\end{table}

Figure~\ref{S-fig:correlations}(a) shows the strong correlations among the raw $D_y$ monomials. Products of the same unit-spin components can carry redundant information about orientation. Most of this structure is removed in the retained PC basis [Fig.~\ref{S-fig:correlations}(b)]. Since PCA is fitted only to the training groups, exact decorrelation is not expected on the held-out set. The weaker test correlations show that the transformation also reduces redundancy outside its fitting sample; they characterize the input representation, not the prediction error.

Figure~\ref{S-fig:xpcs} shows how linear combinations of the raw $D_x$ maps reorganize the allowed angular structure into independent training coordinates. The increasingly intricate patterns supply ways to describe finer orientation dependence while retaining the required mirror-odd character. Figure~\ref{S-fig:ypcs} gives the corresponding nonconstant coordinates for $D_y$; the allowed offset is restored separately. Neither set depicts Berry curvature in momentum space: the axes are moment angles, and the colors are independently normalized basis values. 

PCs are ordered by input variance, rather than polynomial degree or relevance to transport. PCA combines monomials within a single symmetry sector, preserving the required response character when the affine transformation is accounted for. Each PC is assigned the sign that makes its largest-magnitude loading positive. This convention allows consistent coefficient tracking but has no physical significance. The maps describe the angular basis used in main-text Fig.~\ref{fig:predictions}; individual PCs do not identify magnetic phases or microscopic interactions.

\paperfigure[180mm]{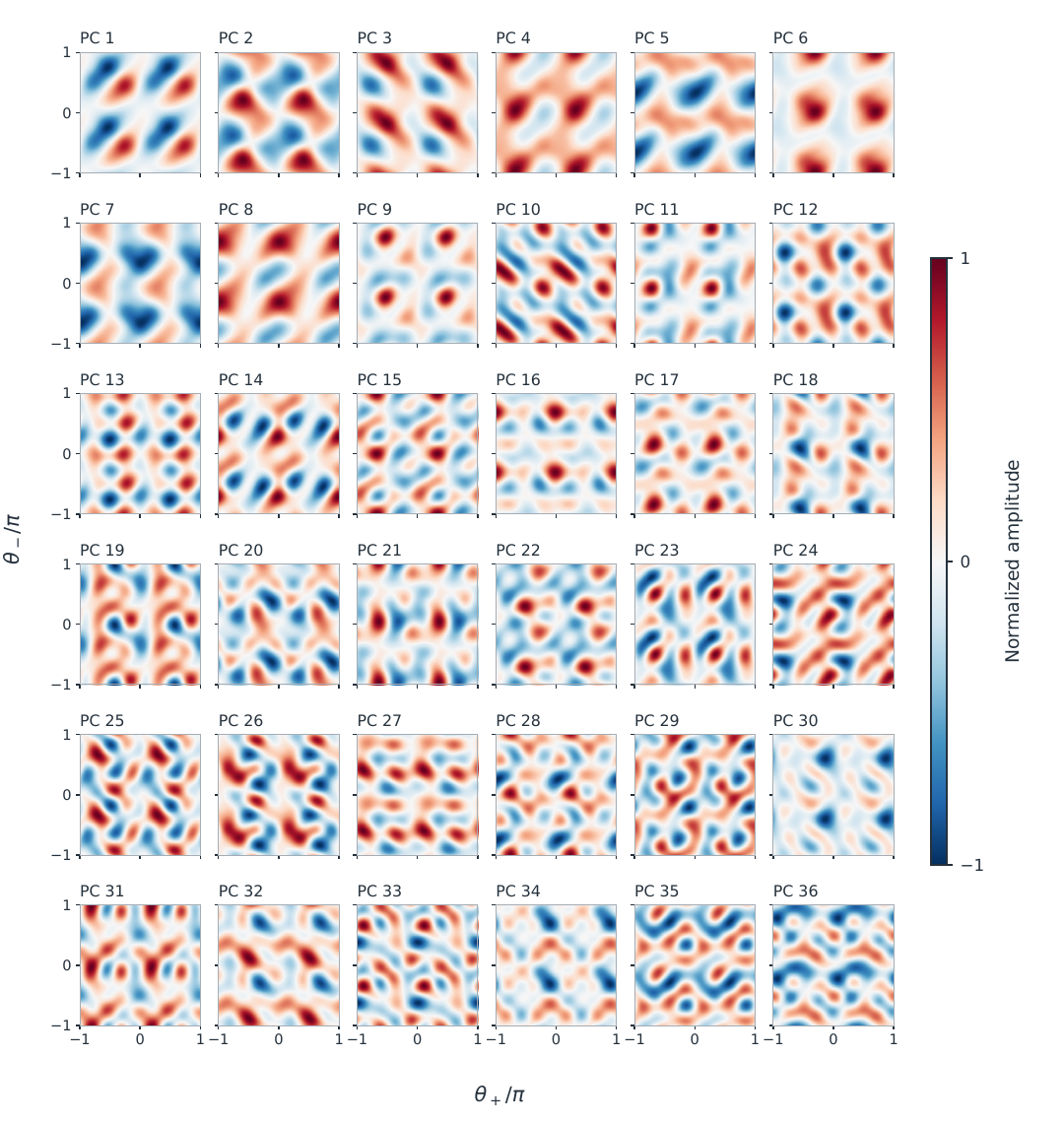}{Angular maps of the first 36 retained $D_x$ PCs. The PCA basis is fitted to the augmented training inputs and contains only the allowed $x$-sector monomials. Each panel is independently normalized. Component numbering starts at one and follows decreasing explained input variance.}{xpcs}

\paperfigure[180mm]{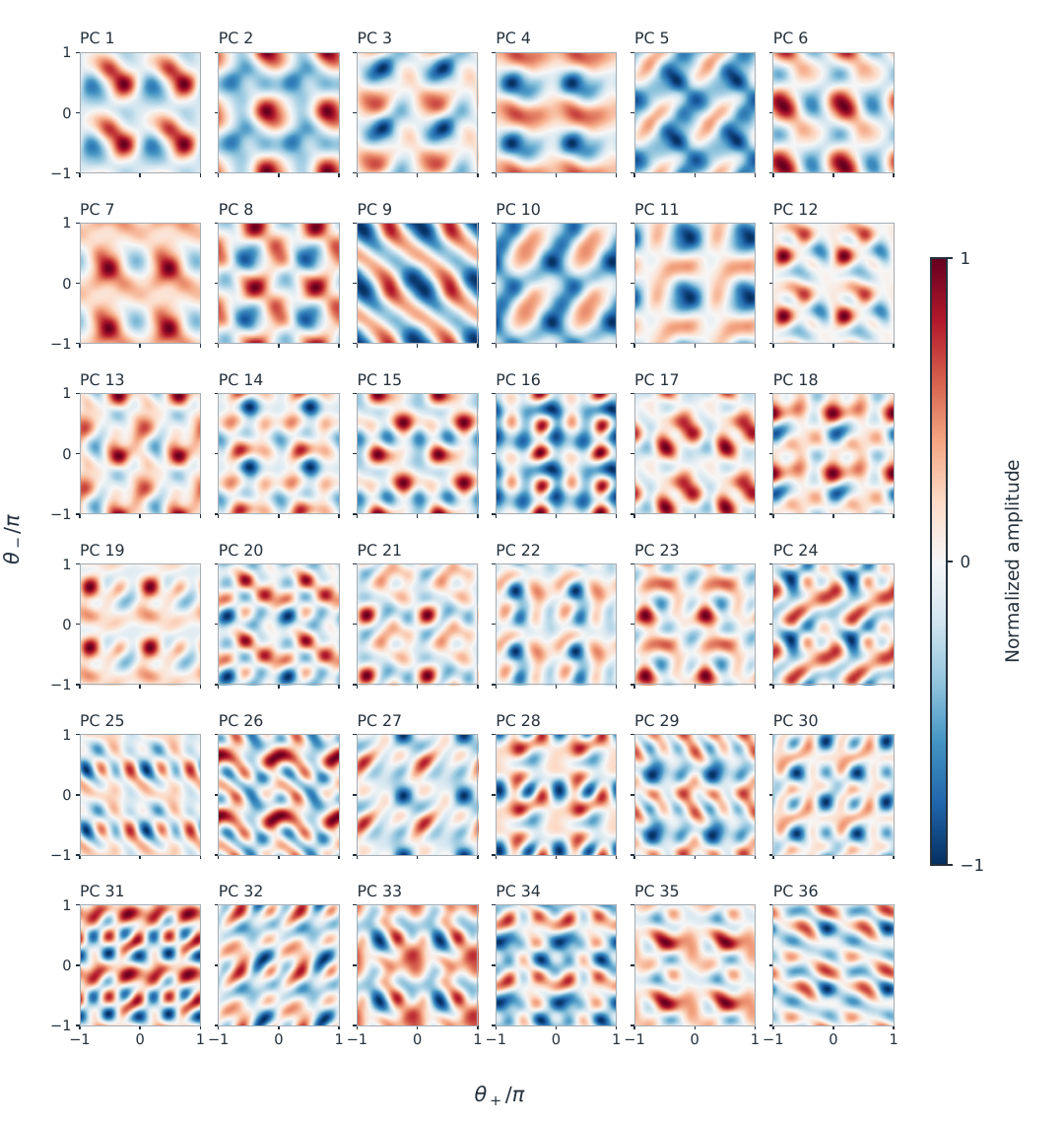}{Angular maps of the first 36 retained $D_y$ PCs, using the same slice as figure~\ref{S-fig:yfeatures}. These are the nonconstant components of the training-only PCA. Their shapes and ordering are determined by the training spin configurations. Each panel is independently normalized.}{ypcs}

\paperfigure[180mm]{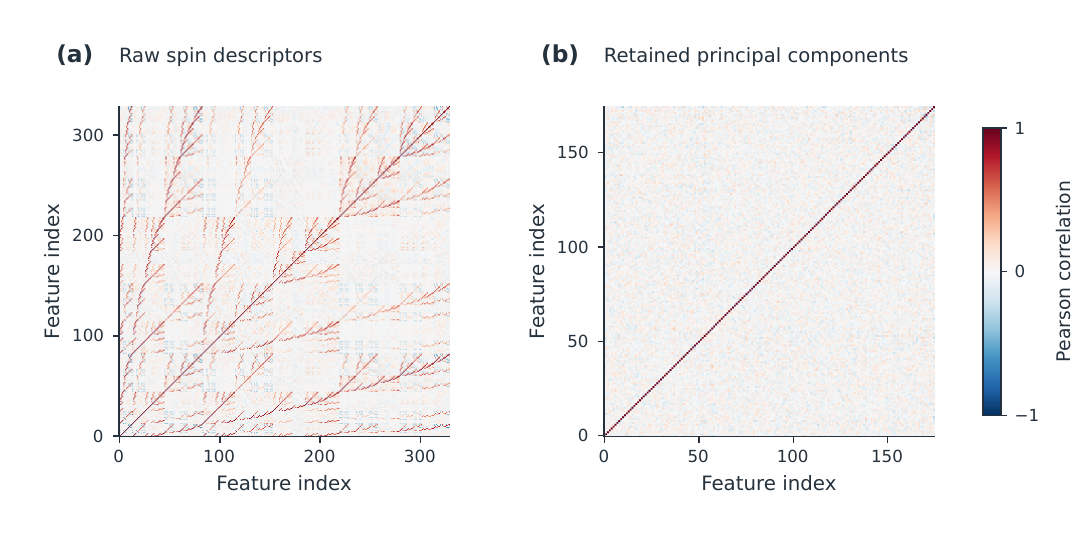}{Pearson correlation matrices on the held-out spin configurations for nonconstant $D_y$ monomials and the retained $D_y$ PCs. The transformation was fitted on training inputs only. Small off-diagonal test correlations are allowed. Null PCs are excluded to avoid displaying correlations determined by numerical roundoff.}{correlations}

\FloatBarrier
\subsection{Input variance and target relevance}\label{S-sec:fresults}
Figure~\ref{S-fig:fscores}(a) compares explained input variance and F-regression scores for every retained $D_y$ component at $\mu_0$. The scatter shows that a component with large input variance need not be strongly correlated with the dipole. The first eight PCs in Fig.~\ref{S-fig:fscores}(b) illustrate this directly: their target scores differ strongly and do not follow the variance ordering. Symmetry supplies allowed angular functions, but the curvature and occupation weights determine their relevance to transport. This is why the pipeline separates an unsupervised numerical-rank reduction from supervised feature selection; a low-variance direction can still contain useful predictive information.

The 30\% and 90\% budgets provide a direct comparison of the sensitivity to retained representation size. They do not constitute a search over all possible sparsity levels. Moreover, an elastic-net coefficient can be small because the associated information is redundant with other selected directions, because of regularization, or because the target depends weakly on it over this sample. 

\paperfigure[180mm]{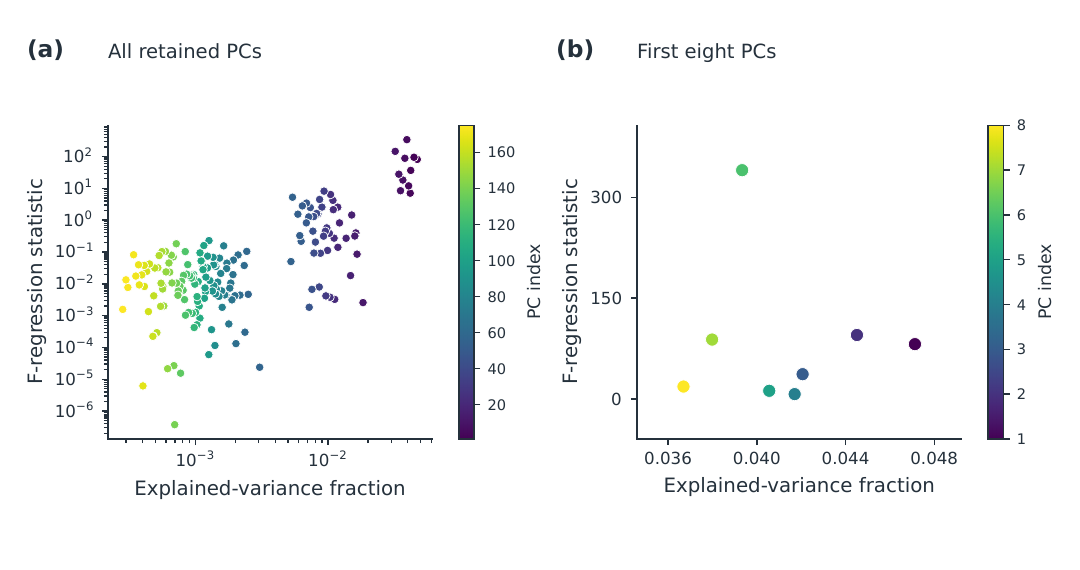}{Explained-variance fraction versus F-regression score for retained $D_y$ components at $\mu_0=0.70\eV$: (a) all retained PCs and (b) the first eight PCs. Color indicates the PC index on the separate scales. Both statistical quantities are obtained from training data. The F statistic is unbounded and does not represent a probability. Input variance and target relevance need not produce the same ranking.}{fscores}

\FloatBarrier
\section{Complete prediction, filling, and energy diagnostics}\label{S-sec:fits}

\subsection{Prediction at fixed chemical potential}
Figure~\ref{S-fig:fixedmu}(a) shows that both feature budgets reproduce the sign and magnitude of $D_y$ across the held-out configurations at fixed chemical potential. The larger budget brings the points closer to the diagonal, and panel (b) resolves the reduced residual spread that is difficult to see on the dipole scale. Increasing from 53 to 158 PCs improves the held-out $R^2$ from $0.995885$ to $0.999518$ and reduces RMSE from $1.0711\times10^{-3}a$ to $3.6642\times10^{-4}a$. 

Figure~\ref{S-fig:fixedmu}(c) displays the narrower central residual distribution of the larger model, but the largest $D_y$ test error remains $3.5654\times10^{-3}a$. A narrow central distribution can therefore coexist with larger errors in its tails. A Gaussian curve is displayed only as a visual guide; neither normality nor a statistical uncertainty interval is inferred from it. Figure~\ref{S-fig:fixedmu}(d) compares the leading standardized weights for the two budgets. Their similar leading contributions are consistent with retaining the main angular dependence while adding corrections, but the bars are representation-dependent coefficients, not measured magnetic coupling constants. Together, panels (a)--(d) explain both the close agreement in main-text Fig.~\ref{fig:predictions}(b) and the remaining limits of that agreement.

The $D_x$ sector shows a comparable benefit from the larger feature budget. Retaining 144 PCs gives $R^2=0.999931$ and RMSE $2.4328\times10^{-4}a$, compared with $R^2=0.997599$ and RMSE $1.4349\times10^{-3}a$ for 48 PCs. Table~\ref{S-tab:fitmetrics} gives both components and both thermodynamic constraints. The main text compares all four larger-budget predictions in a single figure. A regression through degree six is thus highly accurate for interpolation in the specified orientation dataset at $\mu_0$, while retaining identifiable pointwise errors.

\paperfigure[180mm]{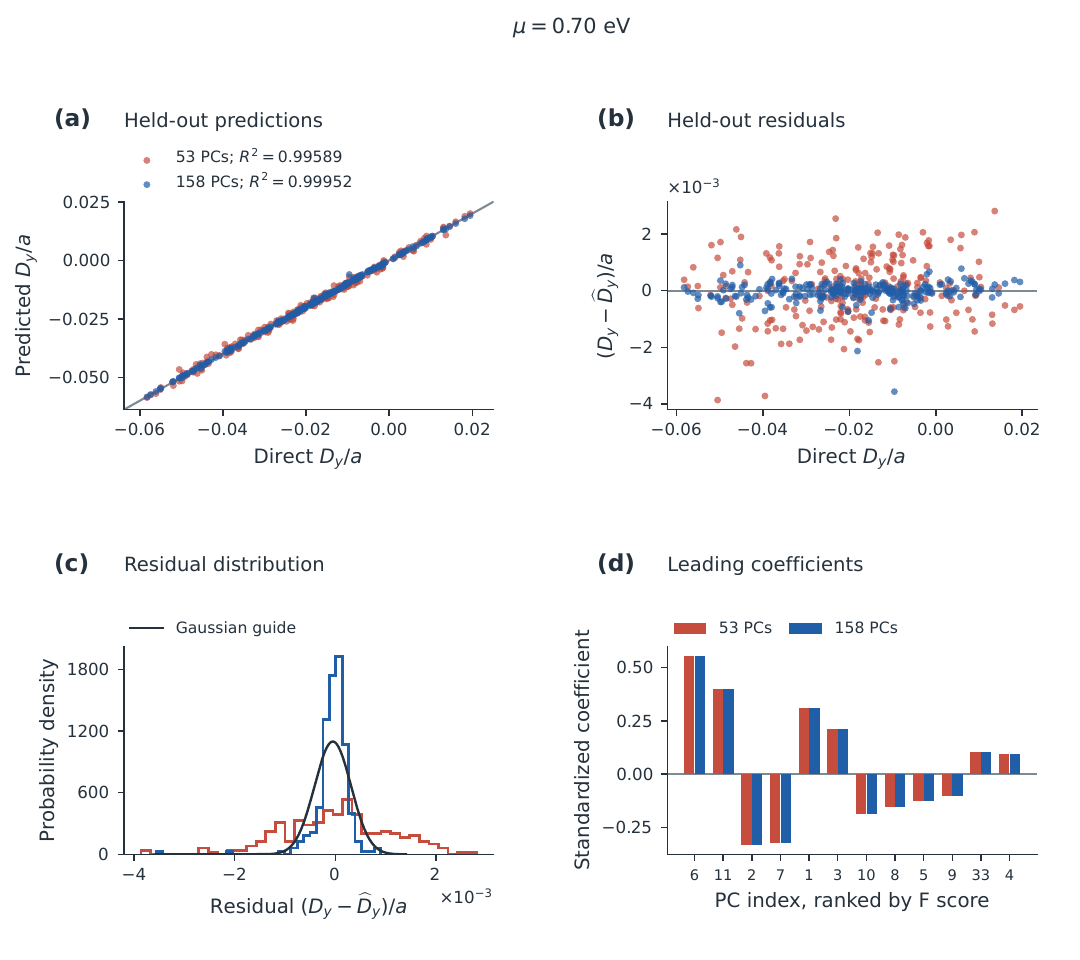}{Fixed-chemical-potential fit for $D_y$ at $\mu_0=0.70\eV$: (a) held-out predictions, (b) residuals, (c) their distribution, and (d) coefficients of leading F-ranked PCs. The smaller and larger models retain 53 and 158 components, respectively. All prediction points are among the 256 held-out configurations. The Gaussian curve is a visual guide, and the coefficient panel uses the standardized fitting basis.}{fixedmu}

\begin{table}[!htbp]
\caption{Held-out prediction statistics for the 256 original test configurations. RMSE and maximum error are in units of $a$.}\label{S-tab:fitmetrics}
\centering\small
\setlength{\tabcolsep}{4pt}
\begin{tabular}{llrrrr}\toprule
Constraint&Target&PCs&$R^2$&RMSE&Max. error\\\midrule
$\mu=0.70\eV$&$D_x$&48&0.997599&$1.4349\!\times\!10^{-3}$&$4.5085\!\times\!10^{-3}$\\
&$D_x$&144&0.999931&$2.4328\!\times\!10^{-4}$&$1.2746\!\times\!10^{-3}$\\
&$D_y$&53&0.995885&$1.0711\!\times\!10^{-3}$&$3.8621\!\times\!10^{-3}$\\
&$D_y$&158&0.999518&$3.6642\!\times\!10^{-4}$&$3.5654\!\times\!10^{-3}$\\\midrule
$n=1.065$&$D_x$&48&0.995087&$1.7997\!\times\!10^{-3}$&$6.3901\!\times\!10^{-3}$\\
&$D_x$&144&0.999756&$4.0092\!\times\!10^{-4}$&$2.1234\!\times\!10^{-3}$\\
&$D_y$&53&0.992251&$1.3112\!\times\!10^{-3}$&$4.6324\!\times\!10^{-3}$\\
&$D_y$&158&0.999077&$4.5250\!\times\!10^{-4}$&$3.8580\!\times\!10^{-3}$\\\bottomrule
\end{tabular}
\end{table}

\FloatBarrier
\subsection{Dependence on chemical potential}
Figure~\ref{S-fig:energy}(a) shows how the weights of ten leading angular components evolve as different parts of the band structure enter the occupation-derivative window. Their changing magnitudes and signs express a redistribution of the orientation-dependent transport contributions. The coefficients are plotted in a common training PCA basis, so this evolution is not caused by redefining the input axes at each chemical potential. Nevertheless, the set selected by F ranking can change with energy. These trajectories summarize a sequence of fits and should not be interpreted as thermodynamic order parameters or evidence of a phase transition at a coefficient zero.

Figure~\ref{S-fig:energy}(b) compares the direct and predicted dipoles for the four smallest held-out IDs, chosen independently of their residuals. Their different energy dependences show that moving the occupation window does not simply multiply every orientation's response by one common factor. Direct and predicted curves are close over much of the interval, with visible deviations in some examples toward the upper end. Table~\ref{S-tab:energy} reports the corresponding statistics over the full test set. At $\mu=0.40\eV$, $R^2$ is $0.999865$ for $D_x$ and $0.999819$ for $D_y$. By $0.90\eV$, it decreases to $0.998100$ and $0.965260$, respectively. The $D_y$ RMSE at this upper endpoint is $5.7499\times10^{-3}a$, with a maximum error of $4.7200\times10^{-2}a$.

\paperfigure[180mm]{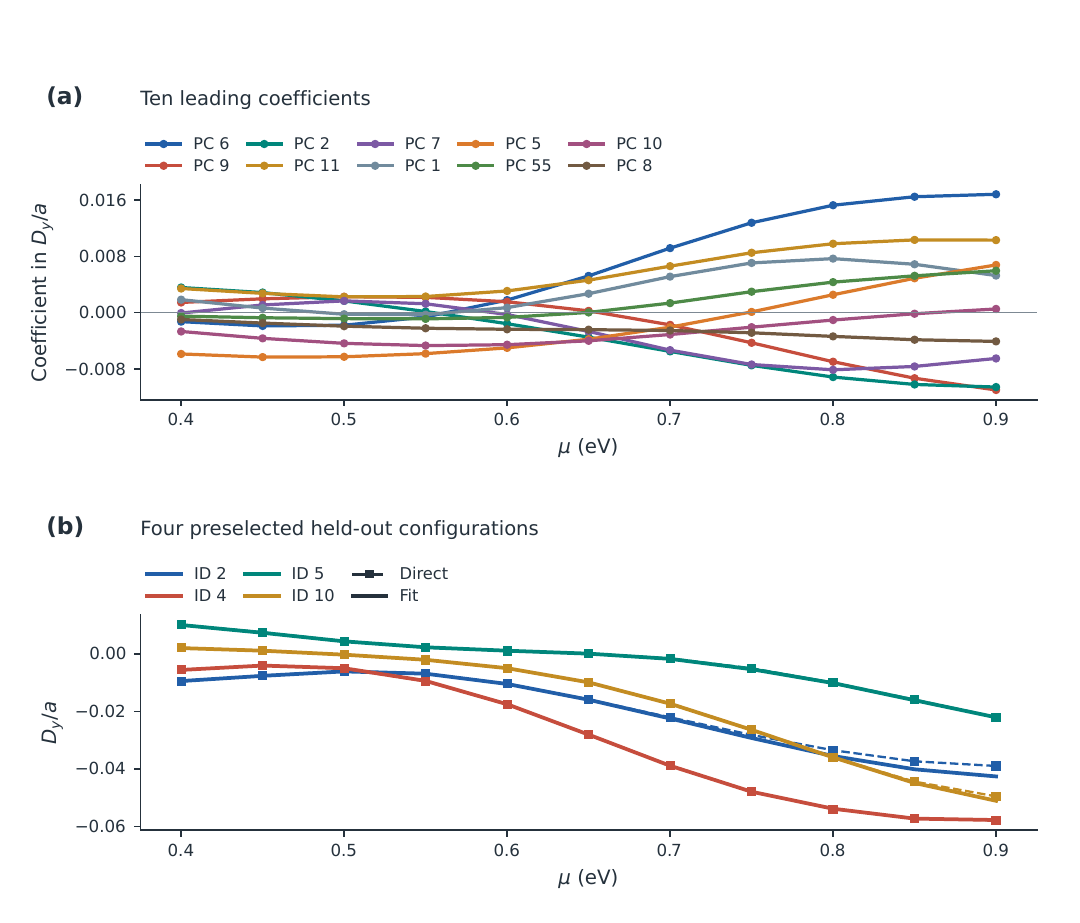}{Energy dependence of the $D_y$ regression. The upper panel shows ten leading coefficient trajectories in the common training PCA basis. The lower panel compares direct labels and predictions for four preselected held-out configurations. Hyperparameters selected at $\mu_0$ are reused; target-based selection and fitting use training labels at each energy. The curves illustrate the fits but do not replace the all-test statistics in table~\ref{S-tab:energy}.}{energy}

\FloatBarrier
\subsection{Filling dependence and fixed-filling regression}
Changing spin orientation modifies band energies as well as eigenvectors. At a common chemical potential, the electronic filling consequently varies between $1.057805$ and $1.083676$ electrons per site across the complete sample. Figure~\ref{S-fig:residualfilling}(a) plots the held-out $D_y$ residuals against filling, colored by the magnitude of $\vect M=(\vect s_A+\vect s_B)/2$. Panel (b) displays the same points colored by the direct dipole. The horizontal spread records the density change caused by rotating the moments at fixed chemical potential, while the vertical spread measures the fit error. The two color scales help examine associations with net moment and dipole magnitude. They do not establish that filling changes cause the residuals, because filling and dipole both depend on the same spin-dependent bands. The separate fixed-filling calculation below tests the effect of imposing a density constraint directly.

\paperfigure[180mm]{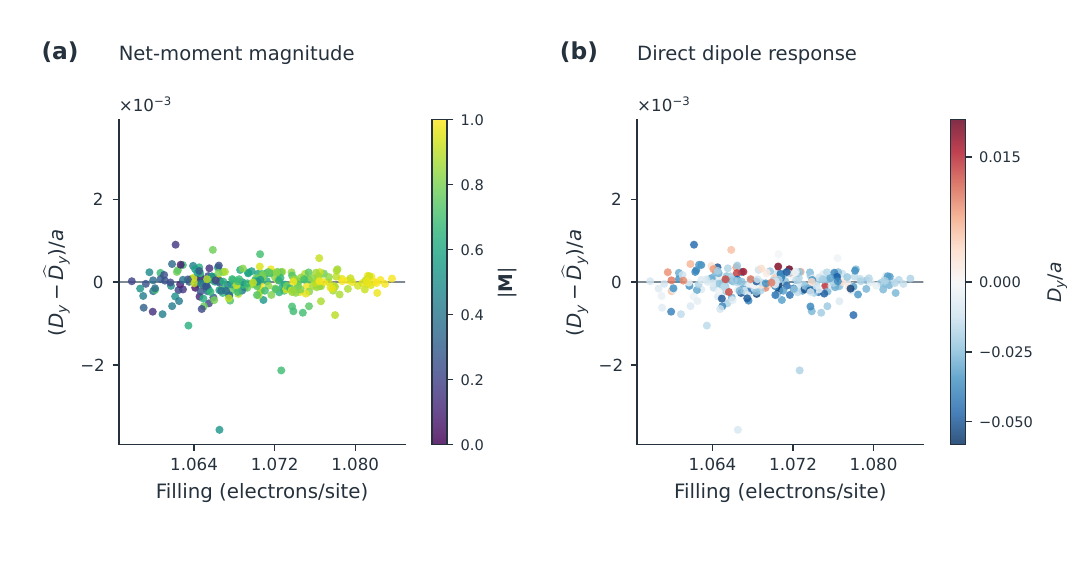}{Fixed-$\mu$ residuals of the larger $D_y$ model versus electronic filling. Color denotes $|\vect M|$ in the left panel and direct $D_y/a$ in the right panel. All points are held out. The plots examine correlations among calculated quantities; no causal conclusion about the origin of the residual is drawn from color or slope alone.}{residualfilling}

A separate target is therefore calculated at fixed filling $n_0=1.065$. Solving the occupation constraint produces chemical potentials from $0.624962$ to $0.728639\eV$ over the complete sample. The maximum final deviation from the requested filling is $6.95\times10^{-14}$ electrons per site within the computed quadrature. This root-solving residual should not be confused with a rigorous error bound on the continuum density integral.

Figure~\ref{S-fig:fixedn}(a) shows that the fixed-filling $D_y$ target remains accurately predictable for the held-out orientations. Panels (b) and (c) expose the residual spread and tails, while panel (d) compares the leading angular weights for the two budgets. The narrower residual core with 158 PCs again shows the benefit of a richer orientation representation, without implying uniformly small errors for every spin pair. Comparing these panels with Fig.~\ref{S-fig:fixedmu} tests the same type of description for two different carrier constraints; the change of target includes the chemical-potential adjustment for each orientation.

The larger fixed-filling model gives $R^2=0.999077$ and RMSE $4.5250\times10^{-4}a$; the corresponding $D_x$ values are $0.999756$ and $4.0092\times10^{-4}a$. Both components remain accurately predicted, although the scores are lower and the errors larger than at fixed $\mu$. The smaller models show the same trend, consistent with main-text Fig.~\ref{fig:predictions} for this Hamiltonian, temperature, and density. Equation~\eqref{S-eq:Dfeedback} separates the direct orientation response from the chemical-potential feedback: maintaining particle number moves the occupation window through the spin-dependent bands. The fixed-filling response thus includes both the changing band geometry and the chemical-potential adjustment.

\paperfigure[180mm]{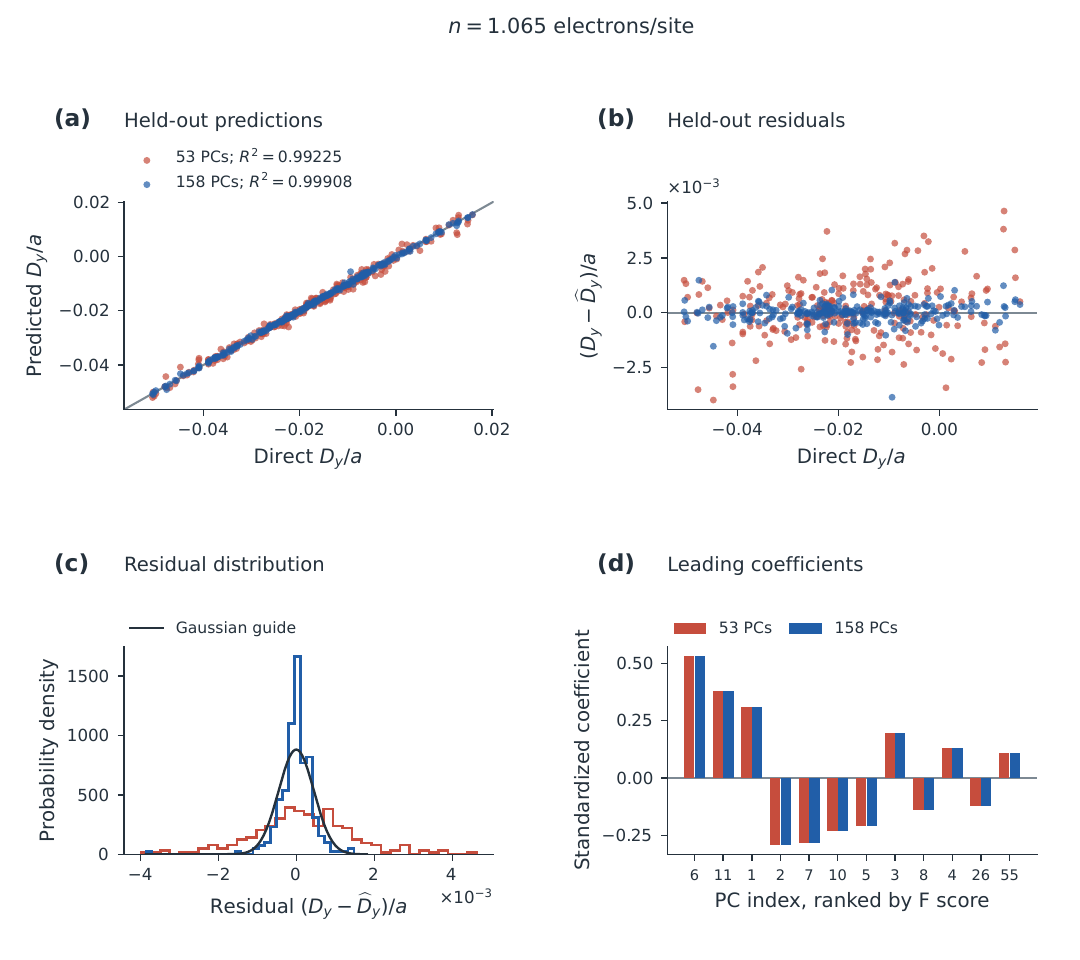}{Fixed-filling fit for $D_y$ at $n_0=1.065$ electrons per site. The chemical potential is solved separately for each spin configuration. Panels and feature budgets match figure~\ref{S-fig:fixedmu}.}{fixedn}

\FloatBarrier
\section{Reproducibility and numerical tables}\label{S-sec:reproducibility}

Table~\ref{S-tab:hyperparameters} lists the selected fitting settings. Reproducing the predictions without refitting requires the configuration split, fitted transformations, selected features, and raw-polynomial coefficients. These calculation artifacts are not included in this manuscript source archive.

\begin{table}[!htbp]
\caption{Cross-validated hyperparameters. Here $\lambda$ is the elastic-net regularization strength, $\rho$ the L1 ratio and $\eta_0$ the initial learning rate in the standardized problem.}\label{S-tab:hyperparameters}
\centering\small
\begin{tabular}{llrrrr}\toprule
Constraint&Target&Budget&$\lambda$&$\rho$&$\eta_0$\\\midrule
Fixed $\mu_0$&$D_x$&30\%&$10^{-5}$&0.15&0.010\\
&$D_x$&90\%&$10^{-4}$&0.85&0.010\\
&$D_y$&30\%&$10^{-5}$&0.15&0.003\\
&$D_y$&90\%&$10^{-3}$&0.15&0.003\\\midrule
Fixed $n_0$&$D_x$&30\%&$10^{-4}$&0.85&0.003\\
&$D_x$&90\%&$10^{-4}$&0.85&0.010\\
&$D_y$&30\%&$10^{-5}$&0.15&0.010\\
&$D_y$&90\%&$10^{-3}$&0.15&0.003\\\bottomrule
\end{tabular}
\end{table}

Table~\ref{S-tab:energy} gives the complete larger-budget sweep. Each energy is a separate fit on the same configuration split.

\begin{table}[!htbp]
\caption{Energy-resolved held-out statistics for the 90\% feature budget. RMSE is in units of $a$. Each row is a separately fitted target evaluated on the same 256 test configurations.}\label{S-tab:energy}
\centering\small
\begin{tabular}{rrrrr}\toprule
$\mu$ (eV)&$R_x^2$&$R_y^2$&RMSE$_x$&RMSE$_y$\\\midrule
0.40&0.999865&0.999819&$1.5520\times10^{-4}$&$1.1684\times10^{-4}$\\
0.45&0.999801&0.999847&$1.7755\times10^{-4}$&$1.1274\times10^{-4}$\\
0.50&0.999878&0.999865&$1.3914\times10^{-4}$&$1.0724\times10^{-4}$\\
0.55&0.999827&0.999821&$1.7391\times10^{-4}$&$1.1875\times10^{-4}$\\
0.60&0.999905&0.999856&$1.4979\times10^{-4}$&$1.0645\times10^{-4}$\\
0.65&0.999953&0.999801&$1.4330\times10^{-4}$&$1.6148\times10^{-4}$\\
0.70&0.999931&0.999518&$2.4328\times10^{-4}$&$3.6642\times10^{-4}$\\
0.75&0.999846&0.998450&$4.6845\times10^{-4}$&$8.8373\times10^{-4}$\\
0.80&0.999669&0.995115&$8.0274\times10^{-4}$&$1.8691\times10^{-3}$\\
0.85&0.999290&0.985261&$1.2915\times10^{-3}$&$3.5498\times10^{-3}$\\
0.90&0.998100&0.965260&$2.2692\times10^{-3}$&$5.7499\times10^{-3}$\\
\bottomrule
\end{tabular}
\end{table}

The portable polynomial agrees with the PCA-space predictions. Its maximum checked covariance residual is $1.4\times10^{-17}a$; the converted current has a transverse-projection residual below $8.1\times10^{-19}$ in normalized units.

\clearpage
\renewcommand{\bibsection}{\section*{Supplemental references}}
\begingroup
\makeatletter\let\label\@gobble\makeatother

\input{supplemental.bbl}
\endgroup